\pdfoutput=1

\documentclass[
preprint, prx,
superscriptaddress,
 amsmath,amssymb,mathtools,
 aps,
]{revtex4-2}

\usepackage{CJKutf8}
\usepackage{graphicx}
\usepackage{siunitx}
\usepackage{chemformula}

\newcommand{\ket}[1]{\left| #1 \right>} 
\newcommand{\Cthirteen}[1]{$\textrm{^{13}C}$} 
\newcommand{\appsection}[1]{\section{\MakeUppercase{#1}}} 

\begin{document}
\begin{CJK}{UTF8}{gbsn}

\title{Detecting spin bath polarization with quantum quench phase shifts of single spins in diamond}

\author{Paul C. Jerger}
\affiliation{Pritzker School of Molecular Engineering, University of Chicago, Chicago, Illinois 60637, USA}

\author{Yu-Xin Wang (王语馨)}
\affiliation{Pritzker School of Molecular Engineering, University of Chicago, Chicago, Illinois 60637, USA}

\author{Mykyta Onizhuk}
\affiliation{Pritzker School of Molecular Engineering, University of Chicago, Chicago, Illinois 60637, USA}
\affiliation{Department of Chemistry, University of Chicago, Chicago, Illinois 60637, USA}

\author{Benjamin S. Soloway}
\affiliation{Pritzker School of Molecular Engineering, University of Chicago, Chicago, Illinois 60637, USA}

\author{Michael T. Solomon}
\affiliation{Pritzker School of Molecular Engineering, University of Chicago, Chicago, Illinois 60637, USA}
\affiliation{Center for Molecular Engineering and Materials Science Division, Argonne National Laboratory, Lemont, Illinois 60439, USA}

\author{Christopher Egerstrom}
\affiliation{Pritzker School of Molecular Engineering, University of Chicago, Chicago, Illinois 60637, USA}
\affiliation{Center for Molecular Engineering and Materials Science Division, Argonne National Laboratory, Lemont, Illinois 60439, USA}

\author{F. Joseph Heremans}
\affiliation{Pritzker School of Molecular Engineering, University of Chicago, Chicago, Illinois 60637, USA}
\affiliation{Center for Molecular Engineering and Materials Science Division, Argonne National Laboratory, Lemont, Illinois 60439, USA}

\author{Giulia Galli}
\affiliation{Pritzker School of Molecular Engineering, University of Chicago, Chicago, Illinois 60637, USA}
\affiliation{Department of Chemistry, University of Chicago, Chicago, Illinois 60637, USA}
\affiliation{Center for Molecular Engineering and Materials Science Division, Argonne National Laboratory, Lemont, Illinois 60439, USA}

\author{Aashish A. Clerk}
\affiliation{Pritzker School of Molecular Engineering, University of Chicago, Chicago, Illinois 60637, USA}
\affiliation{Center for Molecular Engineering and Materials Science Division, Argonne National Laboratory, Lemont, Illinois 60439, USA}

\author{David D. Awschalom}
\email[]{awsch@uchicago.edu}
\affiliation{Pritzker School of Molecular Engineering, University of Chicago, Chicago, Illinois 60637, USA}
\affiliation{Center for Molecular Engineering and Materials Science Division, Argonne National Laboratory, Lemont, Illinois 60439, USA}
\affiliation{Department of Physics, University of Chicago, Chicago, Illinois 60637, USA}

\date{\today}

\begin{abstract}
Single-qubit sensing protocols can be used to measure qubit-bath coupling parameters. However, for sufficiently large coupling, the sensing protocol itself perturbs the bath, which is predicted to result in a characteristic response in the sensing measurements. Here, we observe this bath perturbation, also known as a quantum quench, by preparing the nuclear spin bath of a nitrogen-vacancy (NV) center in polarized initial states and performing phase-resolved spin echo measurements on the NV electron spin. These measurements reveal a time-dependent phase determined by the initial state of the bath. We derive the relationship between sensor phase and Gaussian spin bath polarization, and apply it to reconstruct both the axial and transverse polarization components. Using this insight, we optimize the transfer efficiency of our dynamic nuclear polarization sequence. This technique for directly measuring bath polarization may assist in preparing high-fidelity quantum memory states, improving nanoscale NMR methods, and investigating non-Gaussian quantum baths.
\end{abstract}

\maketitle
\end{CJK}

\section{\label{sec:introduction}Introduction}
A central paradigm of quantum sensing is to probe a target system with a well-understood single qubit. Through the interaction terms, measurements of the qubit reveal details of the target system, which is often chosen to be the qubit’s native environment. Despite the simplicity of a two-level system, these measurements can access a wealth of sensing information, including Hamiltonian parameters~\cite{degen2017quantum}, noise spectra~\cite{Clerk_RevModPhys2010,young2012qubits,Szankowski_JoP2017}, and the presence of entanglement~\cite{roszak2019detect,rzepkowski2021scheme,zhan2021experimental}. Single-qubit sensing has been performed with many common qubit platforms; therefore, advances in sensing protocols have widespread relevance. For instance, echo-based spectroscopy has been employed to characterize the noisy environments of trapped atoms~\cite{biercuk2009experimental}, defect centers~\cite{hernandez2021quantum}, superconducting circuits~\cite{bylander2011noise}, and many other qubits~\cite{almog2011direct,alvarez2011measuring,chan2018assessment,fu2021molecular}.

Echo-based spectroscopy techniques are among the most versatile and widely-used single-qubit sensing methods. In these protocols, qubit control pulses are used to tailor the sensor-target interaction and measure a quantity of interest -- for example, the noise spectral density of the bath within a narrow frequency band~\cite{Szankowski_JoP2017}. While these techniques are powerful and ubiquitous, comparatively little attention has been given to the impact of the sensing protocol on the bath. Naturally, the bath Hamiltonian depends on the qubit state via the qubit-bath coupling terms, but this influence is typically assumed to be negligible, either by treating the bath as a classical noise source~\cite{degen2017quantum,biercuk2011dynamical} or by neglecting the coupling terms when calculating bath dynamics. This assumption is not always justified, and identifying and making predictions for cases where it is violated is an ongoing area of research~\cite{hernandez2018noise,do2019experimental,szankowski2020noise}.

Recent theoretical work explored the impact of a sudden change in the bath Hamiltonian – known as a quantum quench – on the evolution of a sensor qubit during spectroscopy measurements~\cite{Wang_NatComm2021}. This quantum quench is induced by the qubit state rotation at the start of the measurement, and the distinct bath dynamics under the altered Hamiltonian are shown to influence the qubit's final state at the end of the measurement. Specifically, in a pure dephasing environment, Ref.~\cite{Wang_NatComm2021} calculated a phase shift on the final state which depends on the response function of the bath. This quench phase shift (QPS), therefore, encodes valuable environmental information, but has yet to be experimentally observed. Formally, the QPS is of the same order in the qubit-bath coupling as the noise-induced dephasing created by the bath (i.e.,~the quantity that is usually sensed). However, the noise term still dominates when the bath is high temperature, since the noise is then much larger than the susceptibilities which underlie the QPS. Refs.~\cite{PazSilva_PRA2017} and \cite{Kwiatkowski_PhysRevB2020} discuss a related phase shift as the consequence of an anomalous form of qubit-bath coupling, with the latter calculating a phase shift on a nitrogen-vacancy~(NV) center during a spin-echo measurement and predicting it to be linearly proportional to the surrounding nuclear spin bath polarization. These theoretical investigations suggest the NV center and its bath of $^{13}$C nuclear spins as a natural system for detecting and understanding the QPS.

\begin{figure}[t]
    \centering
    \includegraphics[width=80mm]{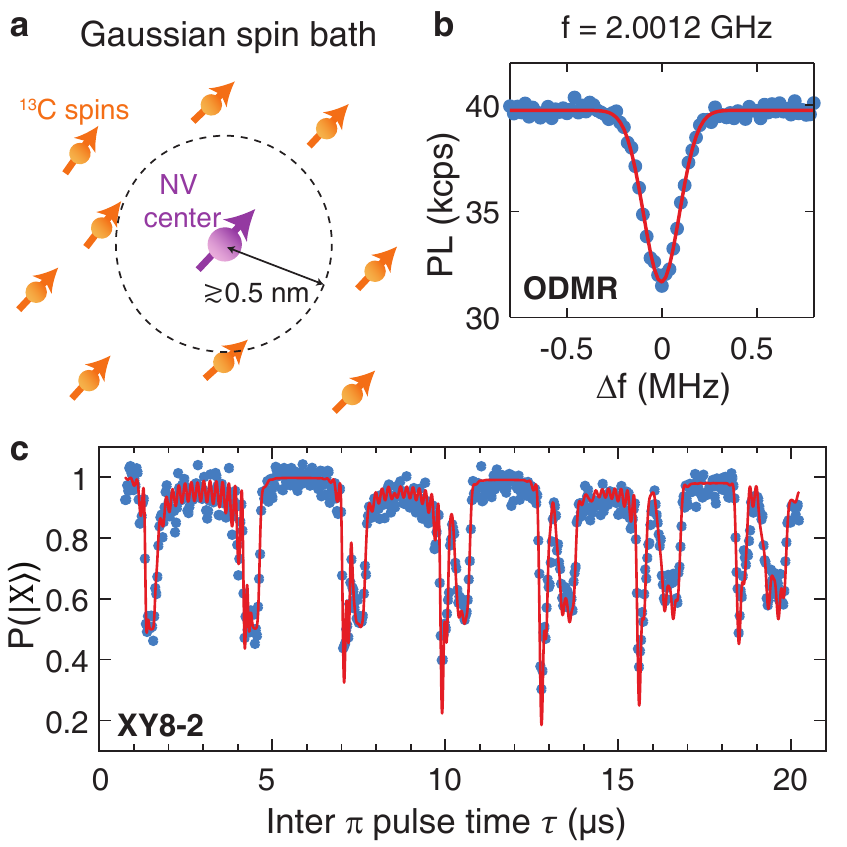}
    \caption{(a) The $^{13}$C spin bath surrounding a nitrogen-vacancy (NV) center in diamond can act as a Gaussian bath if no spins are closer than $\approx$ \SI{0.5}{\nano\meter} to the NV center. By polarizing the bath spins, a polarization-dependent phase appears in NV spin echo measurements, which we observe and analyze under a Gaussian framework. (b) Optically detected magnetic resonance of the NV center corresponding to the $m_I=1$ nitrogen nuclear spin state. The FWHM is $2\pi \times$ \SI{221\pm 6}{\kilo\hertz}, and the single resonance indicates an absence of strongly coupled spins. PL: photoluminescence. (c) Coherence revivals for NV A during an XY8-2 sequence (16 pulses). Fits to the data are used to extract nearby spin coupling parameters, collected in Table 1, and confirm the environment is sufficiently Gaussian.}
    \label{fig:Fig1}
\end{figure}

Here, we present experimental observations of the QPS on single NV centers in diamond. To induce nonzero QPS, we first polarize the bath of $^{13}$C nuclear spins, as in Fig.~\ref{fig:Fig1}a. Spin-echo measurements reveal a QPS with a strong dependence on pulse sequence parameters. To interpret our measurements, we extend the established theory of QPS for a Gaussian spin bath with axial $(\hat{I}_z)$ polarization to encompass transverse $(\hat{I}_x, \hat{I}_y)$ polarization. We show how the phase shift contributions from axial and transverse polarization can be distinguished due to their different physical origins, which allows reconstruction of the collective nuclear spin bath polarization along both axes. These results show how the QPS directly quantifies the NV's local bath polarization, as opposed to existing techniques which provide indirect or only relative polarization information~\cite{Scheuer_PhysRevB2017,Glenn_Nature2018,Bucher_PhysRevX2020}. More broadly, we show that QPS measurements provide access to the bath density matrix and response function, establishing it as a useful technique for investigating the quantum properties of the environment in a variety of single-qubit platforms.

\section{\label{sec:QuenchPhaseModel}Origin of the NV Center Quench Phase Shift}
First, we describe how a quantum quench phase arises in NV center spin echo experiments. The NV center has a spin-1 electronic ground state, and an external magnetic field is applied along its quantization axis. The NV's $^{14}$N nucleus is also spin-1, but is polarized when the applied field is well-aligned~\cite{Busaite_PhysRevB2020}, as in this work, and does not affect the electron spin dynamics~\cite{supplement}. In natural isotopic samples, each NV is also surrounded by a 1.1\%-abundant bath of spin-$\frac{1}{2}$ $^{13}$C nuclear spins. The relevant combined NV-bath spin Hamiltonian can be written in terms of NV electron spin $\hat S_\alpha$ and nuclear spin $\hat I_{\alpha,j}$ ($\alpha = x,y,z$) operators as 
\begin{align}
	\hat H &= \hat H_{NV} + \hat H_{bath} + \hat H_{int}, 
  \label{eq:NVbathH}
  \\
	\hat H_{NV} &=D \hat{S}_z^2 + \gamma_e 
 B_0 \hat{S}_z, \\
\hat H_{bath} &= \gamma_n B_0\sum_j  \hat{I}_{z,j}, 
\label{eq:H0.bath}
\\
\hat H_{int} &= \sum_j \hat{\mathbf{S}}\cdot A_{j}\cdot \hat{\mathbf{I}} _{j}, \label{eq:NVH}
\end{align}
where $D=2\pi\times\SI{2.87}{\giga\hertz}$ is the zero-field splitting, $\gamma_e$ and $\gamma_n$ are the electron and nuclear gyromagnetic ratios, and $A_{j} $ is the hyperfine tensor for the $j$th nuclear spin. Nuclear-nuclear interactions are relatively weak $(\propto \gamma_n^2)$ and can be neglected on the timescales of our experiments. In this work, the electron level splitting dominates all energy scales, so the secular approximation can be applied to recast the hyperfine terms into parallel (axial) and perpendicular (transverse) terms for each nuclear spin as
\begin{equation}
    \hat H_{int} = 
    \hat{S}_z 
    \left (\sum_{j} A_{\parallel,j} \hat{I}_{z,j} + A_{\perp,j}  \hat{I}_{x,j}
    \right ).
    \label{eq:hyperfine}
\end{equation}
From the perspective of bath spins, $\hat H_{int}$ is often described as the NV-state-dependent hyperfine field. With $\hat H_{int}\propto \hat S_z$, the system is well approximated by pure dephasing models. To simplify the problem further, we focus on nuclear spin environments that can be well approximated as Gaussian baths. This limit is best satisfied for baths with many spins, each weakly coupled to the central NV spin. For a given NV center and $^{13}$C distribution, the Gaussian approximation holds for a time $ \tau \lesssim \left(\max_j |A_{\parallel(\perp),j}| \right)^{-1}$. Operating in a pure-dephasing, Gaussian bath regime enables the application of analytic QPS calculations for our system.

Eq.~\eqref{eq:hyperfine} makes clear that an NV spin echo sequence produces an effective quantum quench on the nuclear spin environment. At initialization and as long as the NV remains in $\ket{0}$, $\hat H_{int}$ vanishes, and the bath evolves purely under $\hat H_{bath}$. However, when the NV is rotated to a superposition state, $\hat H_{int}$ once again contributes to bath dynamics. While the consequences of Eq.~\eqref{eq:hyperfine} have been studied and used for nuclear spin sensing and control, the quench dynamics which follow from the sudden activation of $\hat H_{int}$ have only recently been explored. As shown in Ref.~\cite{Wang_NatComm2021}, a quench will occur whenever $ [\hat \rho_{b,i},\hat H_{bath}+
\text{Tr} _{NV} (\hat H_{int}) /2  ]\ne 0$, with initial bath density matrix $\hat \rho_{b,i}$ and $\text{Tr} _{NV}$ denoting partial trace over the NV degrees of freedom. The principal consequence of this quench for spin echo measurements is an additional phase on the final state. Conceptually, this phase arises because the change in bath dynamics conditional on the qubit state creates a net average field from the bath. Concretely, in the NV-bath system, any bath spins initially oriented along the $z$ axis are stationary while the NV state is $\ket{0}$. When the NV state changes, the bath spins begin precessing around a tilted axis due to the $A_{\perp,j} \hat{S}_z \hat{I}_{x,j}$ terms. The precessing bath spins can then induce phase accumulation on an NV superposition state.

Spin echo spectroscopy experiments typically investigate a system by mapping out its coherence function. We can rewrite a generic coherence function $W(\tau)$ as
\begin{align}
    W(\tau) &\equiv
    \frac{\langle\hat{\sigma}_{-}(\tau)\rangle}{\langle\hat{\sigma}_{-}(0)\rangle} \equiv \langle X\rangle -i\langle Y\rangle \nonumber \\
    &= e^{-\chi(\tau)-i\Phi(\tau)}. \label{eq:coherence}
\end{align}
Here, $\hat \sigma_{-} \equiv (\hat \sigma_{x} - i \hat \sigma_{y} )/2 $ denotes the spin lowering operator, $X/Y$ are Bloch vector components, and $\chi$ and $\Phi$ parameterize qubit coherence and phase evolution, respectively. The physical processes which determine $\langle X\rangle=e^{-\chi}\cos\Phi$ have been thoroughly examined in previous studies, finding $\chi$ to be a function of the bath noise spectral density and the filter function of the echo sequence. In the case where the NV undergoes a single (Hahn) spin echo in the presence of a Gaussian spin bath described by Eq.~\eqref{eq:NVbathH}, this has a simple analytic formula,
\begin{equation}\label{eq:chidef}
    \chi (\tau) = 2\epsilon \sin^4 \left( \frac{\omega_L \tau}{4} \right), 
\end{equation}
where $\omega_L$ is the Larmor frequency of the bath spins. The strength of the qubit-bath coupling is parameterized with the dimensionless quantity
\begin{equation}\label{eq:epsilondef}
    \epsilon \equiv \frac{\sum_{j} A_{\perp,j} ^2}{\omega_L^2}.
\end{equation}
Note that $\chi$ is independent of the bath state. The corresponding oscillations in $\langle X\rangle$ have been observed in numerous experiments. In contrast, $\Phi$ is usually found to vanish, reflected in an absence of signal when measuring $\langle Y\rangle=e^{-\chi}\sin\Phi$. We show that nonzero $\Phi$ can provide extensive information on the bath state.

We derive two contributions to $\Phi$, one being the QPS, and both attributable to bath spin polarization. The QPS, denoted $\Phi_q$, can be derived using a linear response approach~\cite{supplement} to be
\begin{equation}
{\label{Eq:PhiQPS}}
    \Phi_{q}(\tau) = \bar{p}_z \epsilon \sin^2{\left(\frac{\omega_L \tau}{4}\right)} \sin{\left(\frac{\omega_L \tau}{2}\right)}.
\end{equation}
Here, $\bar{p}_z$ is the coupling-weighted axial polarization of the bath, 
\begin{equation}
    \bar{p}_z \equiv 
    \frac{\sum_{j} p_{z,j} A_{\perp,j}^2}
    {\sum_{j} A_{\perp,j}^2},
\end{equation}
where $p_{z,j}\in [-1,1]$ is the axial polarization of the $j$th nuclear spin. If $\omega_L$ is known or the $\tau$-dependence is characterized, $\Phi_q$ is determined by only $\bar{p}_z$ and $\epsilon$. Since $\epsilon$ can be characterized independently via Eq.~\eqref{eq:chidef}, measuring $\Phi_q$ provides a direct readout of the axial spin polarization in the local bath.

\renewcommand{\arraystretch}{1.2}
\begin{table}[]
    \centering
    \begin{tabular}{cccccccc}
         \hline
         \hline
         NV\hspace{4pt} & $^{13}$C \# & \hspace{1pt} & $A_\parallel$ (kHz)\hspace{4pt} & $A_\perp$ (kHz) & \hspace{1pt} & $\theta$ (deg) & $r$ (nm)\\
         \hline
           & 1 & \hspace{16pt} & 28.7(3) & 81(1) &\hspace{16pt} & 71 & 0.77 \\
           & 2 & & 29.0(1) & 46.9(9) & & 77 & 0.83 \\
         A & 3 & & -9.8(2) & 27.1(7) & & 35 & 1.27 \\
           & 4 & & 0.3(4) & 23(1) & & 55 & 1.34 \\
           & 5 & & 11.4(4) & 20.2(9) & & 76 & 1.13 \\
         \hline
           & 1 & & -0.1(7)    & 177(1)   &  & 55 & 0.68 \\
           & 2 & & -39.4(7)	& 148(1)	 & & 40 & 0.73 \\
           & 3 & & 87.9(4)    & 122(1)  &  & 102 & 0.58 \\
         B & 4 & & -30.0(2)	  & 80.5(9)	 & & 34 & 0.88 \\
           & 5 & & -16.0(3)     & 72(1) &     & 42 & 0.94 \\
           & 6 & & 51.7(7)    & 58(2)  &    & 100 & 0.71 \\
           & 7 & & -0.1(6)	  & 45(1)	 & & 55 & 1.08 \\
           \hline
           \hline
    \end{tabular}
    \caption{Parallel ($A_\parallel$) and perpendicular ($A_\perp$) $^{13}$C hyperfine parameters obtained from XY8 measurements as in Fig.~\ref{fig:Fig1}(c). Approximate values of $\theta$ (azimuthal angle) and $r$ (NV-nuclear displacement) are calculated assuming pure dipole-dipole interactions.}
    \label{tab:Table1}
\end{table}

A second contribution to $\Phi$ originates from transverse bath spin polarization. Analogous to a phase generated by nearby precessing classical magnetic moments, we label this term $\Phi_m$. Defining $p_{x,j}$ and $p_{y,j}$ in accordance with $p_{z,j}$, $\Phi_m$ enters to first order with hyperfine couplings:
\begin{align}
\label{Eq:phimdef}
    \Phi_m (\tau) = &\frac{2\sin^2\left(\frac{\omega_L \tau}{4}\right)}{\omega_L} \times \\
    &\sum_{j} A_{\perp,j} 
    \left( 
    p_{x,j} \sin\left(\frac{\omega_L \tau}{2}\right)+
    p_{y,j} \cos\left(\frac{\omega_L \tau}{2}\right) \right).
    \nonumber
\end{align}
A full derivation of $\Phi_q$ and $\Phi_m$ is presented in Appendix~\ref{sec:AppADerivations}. Phase shifts related to $\Phi_m$ have been attributed to polarized spin baths in NMR~\cite{marko2013out,sweger2022effect}, ensemble NV~\cite{Glenn_Nature2018,Bucher_PhysRevX2020}, and single NV~\cite{cujia2022parallel} experiments, although these works did not discuss quantitative reconstruction of environmental spin polarization via the measurement. Here, we calculate $\Phi_m$ for a single sensor spin coupled to a Gaussian spin bath. A key step in measuring the QPS is distinguishing $\Phi_q$ from $\Phi_m$, since $\Phi_m$ features leading-order hyperfine terms while $\Phi_q$ is second-order in the couplings, and in general experimental systems exhibit nonzero transverse polarization. As we later show, the distinct physical sources of phase allow us to separate them and extract information about the bath. 

\begin{figure*}[ht]
    \centering
    \includegraphics[width=160mm]{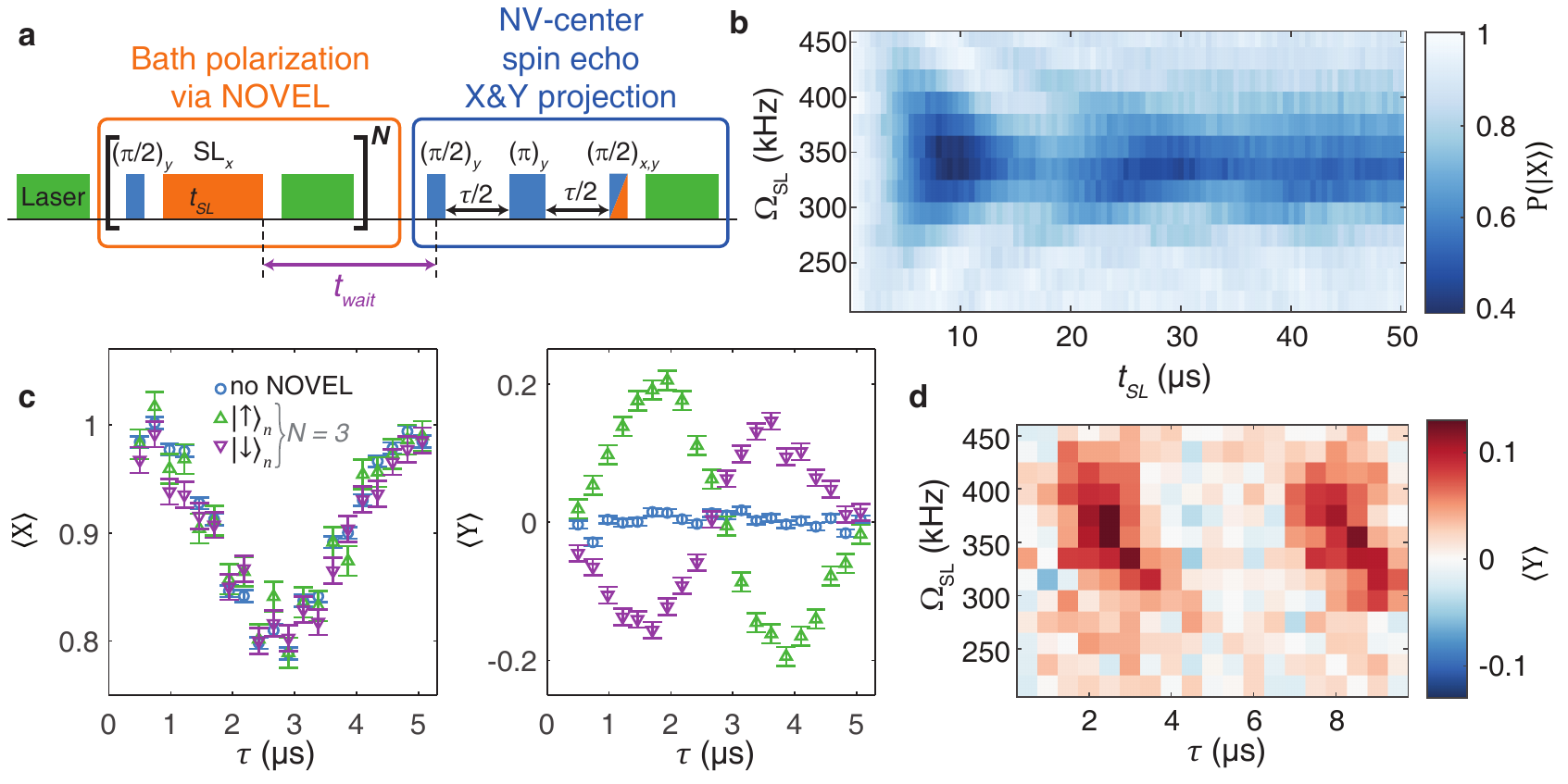}
    \caption{NV center phase shifts appear when the bath is polarized. (a) Initial bath preparation and measurement sequence. $N$ NOVEL repetitions transfer polarization from the NV center to nearby nuclear spins, followed by phase-resolved spin echoes (PSE) to measure both $\left< X\right>$ and $\left< Y\right>$ components of the NV electron spin. (b) NV-bath oscillations during a single NOVEL pulse, after the NV is initialized to $\ket{X}$. Maximal polarization transfer is observed when $\Omega_{SL} = \omega_L =2\pi\times$ \SI{335}{\kilo\hertz}. (c) On resonance, no discernible difference is observed in $\left< X\right>$ (left), but $\left< Y\right>$ (right) shows bath-polarization-dependent oscillations. (d) Using $t_{SL} =$ \SI{4}{\micro\second}, $N = 3$, and preparing the bath into $\ket{\uparrow}$, a phase appears with nuclear Larmor periodicity, but only near resonance. All data shown is taken on NV A.}
    \label{fig:Fig2}
\end{figure*}

\section{\label{sec:Experiments}Single NV Center Experiments}
\subsection{\label{sec:NVGaussianSpinBath}NV Center Characterization and Polarization}
To experimentally investigate the QPS, we study single NV centers in natural isotope abundance IIa diamond at room temperature. We initialize and read out the electronic spin state optically using a \SI{532}{\nano\meter} laser, and apply an external magnetic field $B_0=\SI{310.8}{G}$, aligned to the NV axis within $0.5^{\circ}$. We use a suspended wire coil to apply rf fields for spin operations~\cite{supplement}.

In order to study Gaussian bath dynamics, we identify single NV centers with suitably weak hyperfine couplings. In a perturbative treatment, the Gaussian approximation holds for $ \tau < |A_{\parallel,j}|^{-1}, (A_{\perp,j})^{-1}$ for all spins, and when  $\max_j {|A_{\parallel(\perp),j}|} \ll \omega_L$. Due to the stochastic distribution of $^{13}$C spins around each NV center, some defects have strongly coupled nuclei, which is unfavorable for these criteria. Ideally, a spin-free volume surrounds the NV center, as in Fig.~\ref{fig:Fig1}(a). Ref.~\cite{Kwiatkowski_PhysRevB2020} calculated a minimum spin-free radius of 0.5 nm for the bath to appear Gaussian at moderate magnetic fields. We filter candidate NVs based on narrow optically detected magnetic resonance (ODMR) spectra, as in Fig.~\ref{fig:Fig1}(b), which indicates relatively weak total bath interactions and an absence of individual couplings larger than the linewidth.

We select two NV centers with suitable local spin baths, NV A (FWHM linewidth $221(6)$ kHz) and NV B ($284(14)$ kHz). To fully characterize the respective nuclear spin environments, we apply the XY8 pulse sequence to each NV center to map out hyperfine coupling parameters. Specifically, the XY8-2 sequence with 16 total $\pi$ pulses isolates resonant features in the coherence envelope of the NV center corresponding to the hyperfine interaction with distinct nuclear spins \cite{Taminiau_PhysRevLett2012,Kolkowitz_PhysRevLett2012}. Fig.~\ref{fig:Fig1}(c) shows the coherence data for NV~A, and the best-fit parameters for both NVs are displayed in Table~\ref{tab:Table1}. The data show that both NVs are sufficiently distant from the closest nuclear spins to make the Gaussian approximation reliable for roughly one nuclear Larmor period $T_L=2\pi \omega_L^{-1}$. Using the fit parameters for individual hyperfine couplings, we can compute $\epsilon$ for each NV based on Eq.~\eqref{eq:epsilondef}, and find estimated $\tilde{\epsilon}_A = 0.094$  and $\tilde{\epsilon}_B =0.77$. Using NVs A and B, we investigate the appearance of phase shifts during spin echo experiments. Since the QPS is only predicted to arise with nonzero bath polarization, each measurement involves preparing the initial state of the bath. 

We use the Nuclear Orientation Via Electron spin-Locking (NOVEL) sequence~\cite{Henstra_MolecularPhys2008} to polarize the nearby nuclear spins. The NV is optically polarized to $\ket{0}$ and rotated to $\ket{\pm X}=(\ket{0}\pm \ket{-1})/2$ prior to a spin-locking pulse of duration $t_{SL}$. When the Rabi frequency of the spin-locking pulse, $\Omega_{SL}$, matches $\omega_L$, resonant exchange occurs between the NV and coupled $^{13}$C nuclei. This exchange is shown in Fig.~\ref{fig:Fig2}(b), where the $\ket{X}$ state is measured after a single NOVEL pulse. The electron-nuclear resonance appears at the expected $\gamma_n B_0 = 2\pi \times$ \SI{335}{\kilo\hertz}. By repeating the NOVEL subsequence, polarization accumulates in the bath and persists for much longer than the spin echo timescale of tens of microseconds. The sign of the polarization transfer is determined by the choice of initial NV center state ($\ket{\pm X}$), providing a simple means to invert the bath polarization.

\begin{figure*}[!t]
    \centering
    \includegraphics[width=160mm]{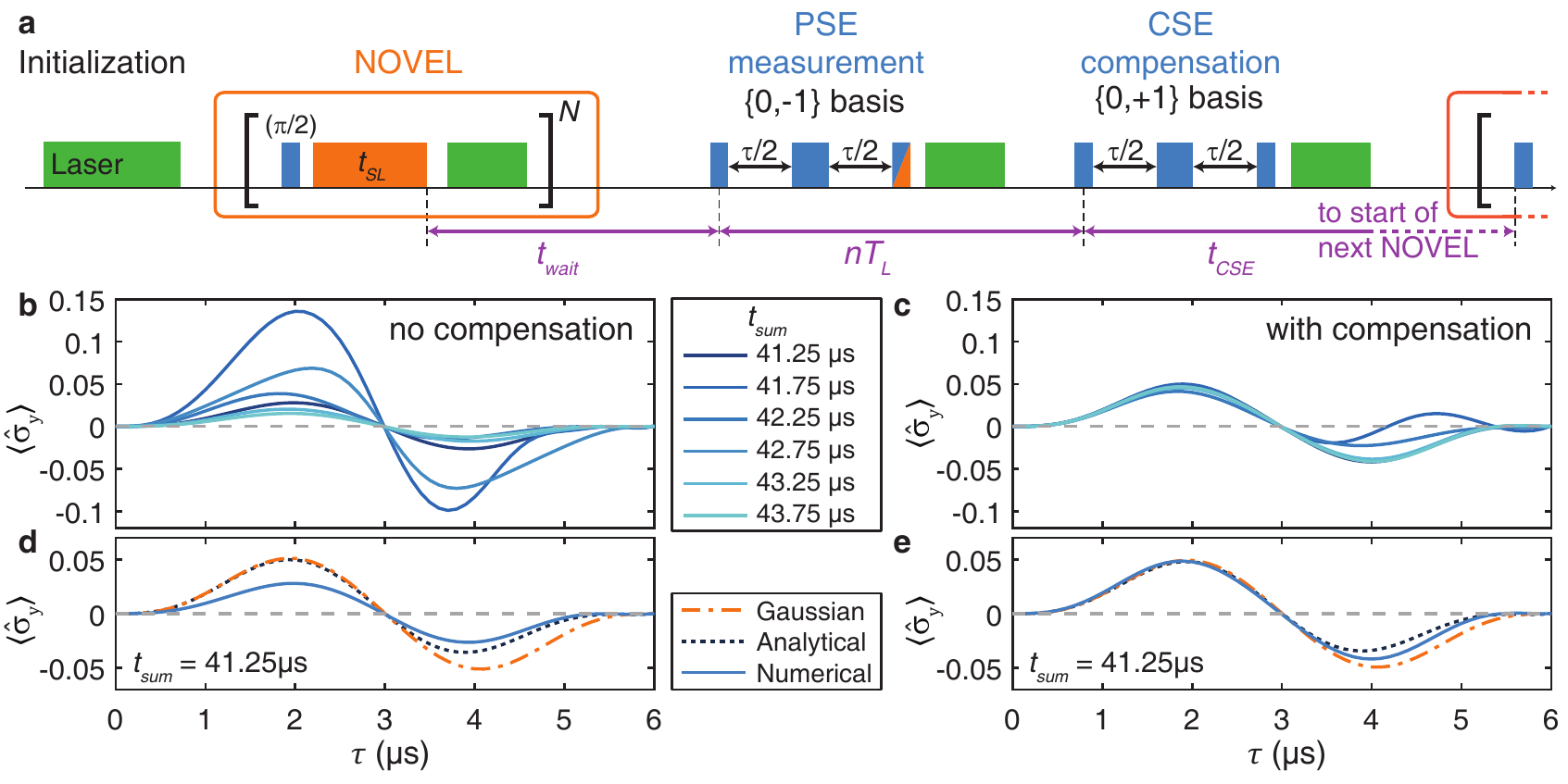}
    \caption{Compensated measurement sequences ensure robust QPS detection. (a) Over many repetitions, the NV-bath coupling during PSE measurement alters the equilibrium bath polarization, but this effect can be minimized by following it with a non-measurement, compensating spin echo (CSE) in the opposite triplet basis. The CSE begins a multiple of the Larmor period ($T_L$) after the start of the PSE. (b) Without CSE pulses, numerical simulations of NV A show the equilibrium bath polarization and resulting PSE signal $\langle \hat \sigma_y \rangle$ are strongly affected by $t_{sum}=t_{wait}+nT_L+t_{CSE}$, obscuring the quench phase shift $\Phi_{q}$. (c) When CSE is included, the bath polarization is robust against changing sequence parameters. Deviations only become significant when the Gaussian approximation falters at longer $\tau$, approximately $\SI{3}{\micro\second}$ for NV A. (d,e) Without CSE pulses, the PSE may not capture $\Phi_q$ accurately, even within the valid Gaussian approximation regime. With CSE pulses, the exact numerical results agree with the Gaussian approximation until the approximation begins to deviate from analytical expectations. For the measurements in this work, $\langle Y\rangle=\langle \hat \sigma_y \rangle$.}
    \label{fig:Fig3}
\end{figure*}

\subsection{\label{sec:SpinEchoPolNucBath}Spin Echoes with Polarized Nuclear Baths}
With a polarized bath, NV spin echo measurements exhibit additional oscillations. The basic measurement framework is illustrated in Fig.~\ref{fig:Fig2}(a), and consists of alternating steps that polarize the bath to a steady state and perform spin echo measurements. After $N$ repetitions of NOVEL re-establish steady-state polarization, we reinitialize the NV to $\ket{0}$ and initiate the spin echo after a delay of $t_{wait}$. Using different phases for the spin echo readout pulse, we measure the $\pm X$ and $\pm Y$ components of the final state to reconstruct the amplitude and phase – a phase-resolved spin echo (PSE). The combined NOVEL+PSE sequence is then repeated $10^6$-$10^7$ times to record average statistics. Conventionally, only $\langle X\rangle$ is measured in an echo experiment, with NV-bath interactions producing coherence oscillations resembling the blue data of Fig.~\ref{fig:Fig2}(c). With a high-temperature bath, $\langle Y\rangle$ provides no additional information. However, when the bath is polarized ($N=3$, green/purple curves), clear oscillations in $\langle Y\rangle$ appear on the timescale of the Larmor period. These phase oscillations follow the sign of the bath polarization (Fig.~\ref{fig:Fig2}(c)) and are correlated with the spin-locking resonance (Fig.~\ref{fig:Fig2}(d)), clearly linking their origin to the polarized bath spins. Note that $\langle X\rangle$ shows no difference regardless of bath polarization; for Gaussian baths, polarization has an imperceptible effect on this projection. Since $\epsilon$ is independent of the bath state, we fit the $\langle X\rangle$ data for each NV to Eq.~\eqref{eq:chidef}, resulting in measured values $\epsilon_{A}=0.110(9)$ and $\epsilon_{B}=0.68(5)$. These values are in good agreement with the estimations $\tilde{\epsilon}_{A}$ and $\tilde{\epsilon}_{B}$ from Sec.~\ref{sec:NVGaussianSpinBath}, increasing confidence in the extracted hyperfine parameters.

Before quantifying and further investigating the spin echo dynamics, we refine the measurement protocol to robustly extract the QPS. Since the combined measurement sequence modifies the spin bath by design, the PSE sequence necessarily alters the bath preparation entering the next repetition of the experiment. This can produce a confounding effect when sweeping a spin echo or NOVEL parameter, as we demonstrate through numerical simulations. In Fig.~\ref{fig:Fig3}, we simulate the exact spin dynamics of NV A and its nearby $^{13}$C spins, using the measured hyperfine parameters, while evolving under the preparation and measurement sequence of Fig.~\ref{fig:Fig3}(a)~\cite{supplement}. The simulated sequence is repeated until the bath reaches a steady state. We calculate $\langle \hat \sigma_y\rangle$, which is equal to $\langle Y\rangle$ for our measurements. Each trace is an average over a range of $t_{wait}$ to reduce the effects of nuclear precession, which isolates the $\Phi_q$ component, as will be addressed in more detail in Section~\ref{sec:PhaseShifts}. In Fig.~\ref{fig:Fig3}(b), we begin by incorporating only the NOVEL and PSE elements, followed by a variable wait period. We define $t_{sum}$ as the interval between the $N$th NOVEL pulse and the first NOVEL pulse of the next repetition. Despite only changing $t_{sum}$, i.e., varying the wait time between repetitions, we note dramatic changes in the behavior of $\langle \hat \sigma_y\rangle$. This indicates the steady-state bath polarization can strongly depend on the measurement parameters in addition to the NOVEL parameters.

\begin{figure*}
    \centering
    \includegraphics[width=160mm]{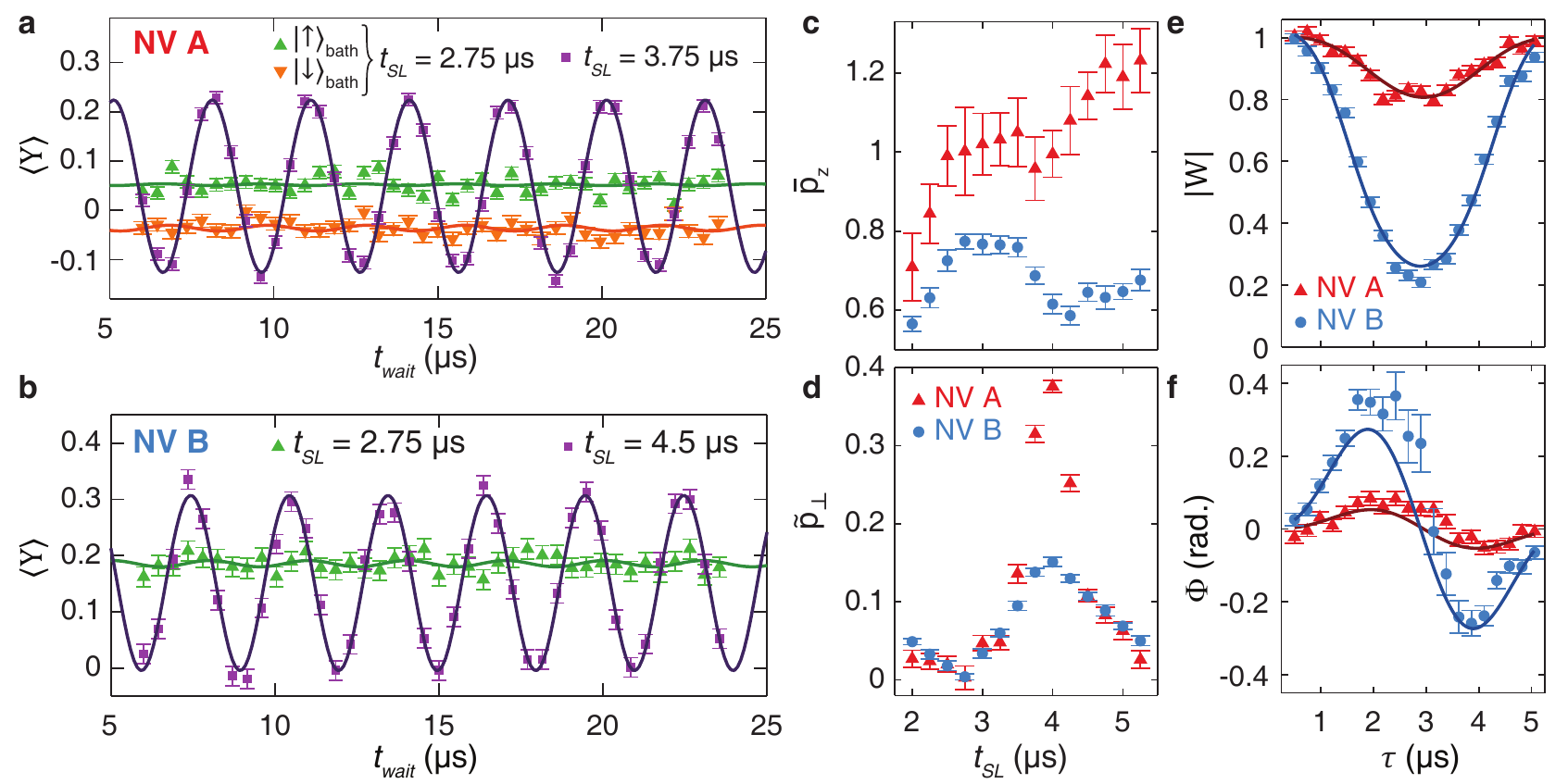}
    \caption{Observation of quench phase shifts and corresponding polarization measurements. (a,b) Phase evolution with $\tau = $\SI{1.5}{\micro\second} on NVs A (a) and B (b) with respect to $t_{wait}$, and for several $t_{SL}$. The oscillations correspond to the precession of transverse nuclear polarization, manifesting in an oscillating $\Phi_m$ as described in Eqs.~\eqref{Eq:phimdef} and \eqref{Eq:Eq_pxpy}. The constant offset is $\Phi_{q} \propto \bar{p}_z$. NV B, with stronger hyperfine couplings, exhibits both larger $\Phi_m$ and $\Phi_{q}$. (c,d) The bath polarization state is optimized by tuning $t_{SL}$. Each point is determined by fitting a sweep of $t_{wait}$, as in (a,b), to the combined Eqs.~\eqref{Eq:PhiQPS}, \eqref{Eq:phimdef}, and \eqref{Eq:Eq_pxpy}. (e,f) The time evolution of the coherence magnitude $|W|$ and $\Phi$, fit to Eq.~\eqref{Eq:PhiQPS}, with $\tilde{p}_\perp$ minimized for each NV using $t_{SL}=$\SI{2.75}{\micro\second}.}
    \label{fig:Fig4}
\end{figure*}

To mitigate this effect, we introduce a second, non-measurement spin echo performed on the $\{0,+1\}$ basis following the PSE in the $\{0,-1\}$ basis, which we refer to as a compensating spin echo (CSE). Because of the symmetry between the $\ket{\pm 1}$ states, the net effect of the CSE is to reverse the PSE's perturbation on the bath state to lowest order. Importantly, for the most effective compensation, the initial pulses of the PSE and CSE are separated by a multiple of $T_L$, so that the bath spins are close to their state at the beginning of the PSE. The benefit of compensation is shown in Fig.~\ref{fig:Fig3}(c), where the simulations are repeated including the CSE. The echo signal is observed to be robust, indicating a bath state which is not sensitive to parameters in the measurement sequence. At large $\tau$, the computed $\langle \hat \sigma_y\rangle$ does eventually exhibit noticeable difference for different $t_{sum}$, but only after the Gaussian approximation begins to break down. Heuristically, the PSE and CSE can each be viewed as performing a `pulse' on the surrounding bath. While the effect of the first such pulse may be complicated in general, the bath can be restored to approximately its initial state by performing an inverse pulse with appropriate timing. A detailed analysis of the CSE's compensating effect is presented in Appendix~\ref{sec:AppC_CSE}, with additional numerics in the Supplementary Materials~\cite{supplement}.

In addition to establishing the robustness of the augmented measurement protocol, we confirm that it accurately quantifies the QPS. Figs.~\ref{fig:Fig3}(d) and \ref{fig:Fig3}(e) show three separate calculations of $\langle \hat \sigma_y\rangle$ for $t_{sum}=\SI{41.25}{\micro\second}$. These calculations each stem from the same simulation of the NV A spin cluster, but incorporate the steady-state bath polarization components $\bar{p}_z$ and $p_\perp = \sqrt{p_x^2+p_y^2}$ (at the start of the PSE) in different ways, as follows:
\begin{enumerate}
    \item[(i)] The Gaussian curve (dashed orange) plots the QPS signal for a Gaussian bath with $\epsilon=\epsilon_A$ and $\bar{p}_z$ only.
    \item[(ii)] The analytical curve (dotted black) plots the exact dynamics for $\bar{p}_z$ only.
    \item[(iii)] The numerical curve (solid blue) plots the exact dynamics including both $\bar{p}_z$ and $p_\perp$. 
\end{enumerate}

Without the CSE (Fig.~\ref{fig:Fig3}(d)), the $\langle \hat \sigma_y \rangle$ signal is biased by $p_\perp$ contributions and differs from $\Phi_q$ by roughly a factor of 2. Including the CSE (Fig.~\ref{fig:Fig3}(e)) eliminates the bias, indicating that $\Phi_q$ can be extracted by averaging over $t_{wait}$. These simulations also confirm that the Gaussian approximation is reliable for $\tau\lesssim \SI{3}{\micro\second}$. In ensuing experiments, we use the full compensated PSE sequence to quantify $\Phi_m$ and $\Phi_q$.

\subsection{\label{sec:PhaseShifts}Phase Shift Measurements}
Using the measurement sequence of Fig.~\ref{fig:Fig3}(a), we observe phase shifts on our single NV centers which correspond to the polarization and precession of the nuclear spin bath. In Fig.~\ref{fig:Fig4}(a) (NV A) and \ref{fig:Fig4}(b) (NV B), we vary $t_{wait}$ while holding all other sequence parameters constant. We choose $\tau=\pi/\omega_{L}=\SI{1.5}{\micro\second}$, which maintains large signal while simplifying Eqs.~\eqref{Eq:PhiQPS} and \eqref{Eq:phimdef} to
\begin{align}
    \Phi_q\left(\tau=\frac{\pi}{\omega_L}\right) &= \frac{\bar{p}_z\epsilon}{2},\label{eq:phiq_tau1.5} \\
    \Phi_m\left(\tau=\frac{\pi}{\omega_L}\right) &= \frac{1}{\omega_L} \sum_j A_{\perp,j}\  p_{x,j}.\label{eq:phim_tau1.5}
\end{align} The resulting oscillations match $\omega_L$ and arise from the precession of the nuclear spin state after the final NOVEL pulse. The NOVEL preparation gives rise to $p_\perp >0$ through two mechanisms: appreciable $A_\parallel$ components relative to $\omega_L$, and spin-locking in the asymmetric $\{\ket{0},\ket{-1}\}$ basis. These sources are discussed in more detail in Appendix~\ref{sec:AppBTransPol}. The latter is particularly easy to overlook for the NV center, since its triplet structure is often reduced to a two-level system to simplify calculations. As described by Eq.~\eqref{Eq:phimdef}, any transverse polarization produces a $\left< Y\right>$ signal. After the final NOVEL pulse, the initial transverse polarization will precess during $t_{wait}$ between $p_x$ and $p_y$:
\begin{equation}
\label{Eq:Eq_pxpy}
\begin{aligned}
    p_{x,j} &= p_{\perp,j}\cos{\left(\omega_L t_{wait} + \varphi_{j}\right)}, \\  
    p_{y,j} &= p_{\perp,j}\sin{\left(\omega_L t_{wait} + \varphi_{j}\right)}.
\end{aligned}
\end{equation}
$\varphi$ is the initial phase of the transverse polarization. Previously, related oscillations have been detected in ensemble experiments \cite{Bucher_PhysRevX2020,Glenn_Nature2018}, but are observed here at the single-NV level and quantified in a Gaussian framework. Combining Eq.~\eqref{eq:phim_tau1.5} with the parameters in Table~\ref{tab:Table1}, and assuming a uniform initial bath polarization via $p_{\perp,j}=\tilde{p}_\perp$, $\varphi_{j}=\varphi$, we can estimate the mean transverse polarization $\tilde{p}_\perp$. By fitting the oscillations, we find $\tilde{p}_\perp =0.32(1)$  for NV A  using $t_{SL}=\SI{3.75}{\micro\second}$, and $0.107(5)$ for NV B using $t_{SL} = \SI{4.5}{\micro\second}$ (purple data sets of Figs.~\ref{fig:Fig4}(a) and \ref{fig:Fig4}(b)). By tuning $t_{SL}$, we minimize $\tilde{p}_\perp$ (Fig. \ref{fig:Fig4}(d)), achieving $\tilde{p}_\perp = 0.003(15)$ for NV A and $0.004(4)$ for NV B. These measurements do not suffice to uniquely determine the transverse polarization of each nearby nuclear spin, but do provide a means of rapidly estimating it with a single quantity – the oscillation amplitude – to adjust for nonideal behavior in polarization sequences. Even without quantifying the bath's hyperfine constants, this metric provides qualitative feedback when optimizing bath preparation sequence parameters.

Within the same PSE data set, we observe the QPS and use it to quantify $\bar{p}_z$. In addition to the oscillating $\Phi_m$ component, we detect a constant phase offset as a function of $t_{wait}$. $\Phi_q$ is independent of $t_{wait}$, since decay of nuclear polarization is negligible on the timescale of $t_{wait}$. The $\Phi_m$ component of $\Phi$ can be canceled by averaging $\left< Y\right>$ over $t_{wait}$ or, equivalently, fitting the $\left< Y\right>$ oscillations to an offset. Thus, both $\Phi_m$ and $\Phi_{q} \propto \bar{p}_z\epsilon$ can be fit simultaneously to $\langle Y\rangle =e^{-\chi}\sin \Phi$, using the previously measured $\epsilon$. For the optimal values of $t_{SL}$ (green data sets), we measure $\bar{p}_{(z,A)} = 1.00(11)$ and $\bar{p}_{(z,B)} = 0.77(2)$.

The observed $\Phi_q$ exhibits the predicted QPS characteristics. $\Phi_{q}$ is sensitive to the direction of bath polarization, as in Fig. \ref{fig:Fig4}a, where the offset inverts along with the bath state. When transverse polarization is present, sweeping $t_{wait}$ is necessary to confirm the sign of the offset, since $|\Phi_m|$ can exceed $|\Phi_{q}|$. Alternatively, with fixed $t_{wait}$, $\tau$ can be swept to map the $\Phi(\tau)$. Both $\langle X\rangle$ and $\langle Y\rangle$ are needed to correctly calculate $|W|\equiv e^{-\chi}$ and $\Phi$. In Figs.~\ref{fig:Fig4}(e) and \ref{fig:Fig4}(f), the fits to $\Phi_q$ for both NVs with minimized $\tilde{p}_\perp$ show that $\Phi_{q}$ closely matches the predicted dynamics of Eqs.~\eqref{eq:chidef} and \eqref{Eq:PhiQPS}.

As with $\tilde{p}_\perp$, we examine the dependence of $\bar{p}_z$ on $t_{SL}$. In Fig.~\ref{fig:Fig4}(c), both NVs exhibit similar trends: $\bar{p}_z$ improves where $t_{SL}\sim T_L$, and is reduced where $\tilde{p}_\perp$ is largest. While predicting $t_{SL}$ dependence in general requires knowledge of the spin bath, PSE measurements offer a simple method to investigate parameter sensitivity without exhaustively characterizing the bath. $\bar{p}_z$ exceeding unity is observed for NV A, and can be attributed to uncertainty in $\epsilon_A$, which is used as a constant for calculating all $\bar{p}_{z,A}$ values. The confidence intervals for NV A are larger than NV B since the QPS is smaller for A, leading to larger fractional uncertainty. We find good agreement between the experimental data and numerical predictions for NV A using the identified nuclear spins~\cite{supplement}.
 
Next, we demonstrate that small $\Phi_{q}$ signals can be increased with additional spin echoes in the PSE sequence. In general, target bath systems may have smaller $\epsilon$ than studied here, and improved signal reduces the  need for extensive averaging. For spin baths, smaller $\epsilon$ may be due to more distant spins $(A_\perp \propto r^{-3})$ or larger magnetic fields $(\omega_L \propto B_0)$, as captured by Eq.~\eqref{eq:epsilondef}. Applying similar linear response analysis to the general case of an $M$-pulse dynamical decoupling sequence of the form $\left[ \frac{\tau}{2} - \pi - \frac{\tau}{2}\right]^M$, the resulting sensor qubit evolution is given by
\begin{align}
& \chi(\tau) = 
\begin{cases}
2\epsilon\frac{\sin^2\left(\frac{M\omega_{L}\tau}{2}\right)
\sin^4\left(\frac{\omega_{L}\tau}{4}\right)}
{\cos^2\left(\frac{\omega_{L}\tau}{2}\right)},\vspace{8pt} & \text{M is even}
,\\
2\epsilon\frac{\cos^2\left(\frac{M\omega_{L}\tau}{2}\right)\sin^4\left(\frac{\omega_{L}\tau}{4}\right)}{\cos^2\left(\frac{\omega_{L}\tau}{2}\right)}, & \text{M is odd},
\end{cases}
\label{eq:multipulseChi}
\\
&\Phi_{q}(\tau)=(-1)^{M-1}
\bar{p}_z\epsilon
\frac{\sin\left(M\omega_{L}\tau\right)\sin^2\left(\frac{\omega_{L}\tau}{4}\right)}
{2\cos\left(\frac{\omega_{L}\tau}{2}\right)}.
\label{eq:multipulsePhi}
\end{align}
 See Appendix~\ref{appsec:qbcoh.13C.CPMG} for the derivation of these expressions.

As $\epsilon \rightarrow 0$, the polarization information in $\Phi_{q}$ coincides with $\langle Y(\tau)\rangle = e^{-\chi(\tau) }\sin(\Phi_{q})$. Maximizing the measurement signal is therefore a balance between the loss of coherence and the accumulation of phase. For Gaussian spin baths described by Eqs.~\eqref{eq:H0.bath} and~\eqref{eq:hyperfine}, the theoretical upper bound of the QPS signal is $|\langle \hat \sigma_y \rangle | \lesssim 0.3 \sqrt{\epsilon}$, which can be saturated when $\epsilon < 1$ and $\chi = \frac{1}{2}$ (see Appendix~\ref{appsec:qbcoh.13C.sy.lim}). At this maximal signal point,
\begin{equation}
M_{opt} \sim \frac{1}{\sqrt{\epsilon}}, \quad
\tau_{opt} = \frac{\pi}{\omega_L \sqrt{\epsilon}}
.
\label{eq:optMultipulse}
\end{equation}

\begin{figure}
    \centering
    \includegraphics[width=80mm]{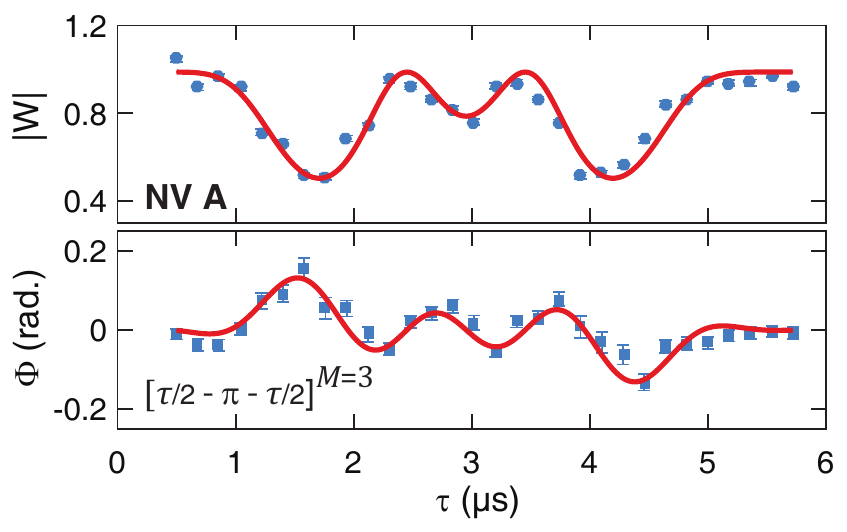}
    \caption{PSE measurements with additional echoes can increase signal. As demonstrated on NV~A, the phase $\Phi$ (bottom) has greater amplitude for $M=3$ echoes than for $M=1$. Correspondingly, the coherence magnitude $|W|$ (top) is diminished at the maximum of $\Phi$, so the signal cannot be increased monotonically with additional pulses. The data are fit to Eqs.~\eqref{eq:multipulseChi} and \eqref{eq:multipulsePhi}, with the optimal $M=3$ determined by Eq.~\eqref{eq:optMultipulse}. Here, $\tau$ is the interval between echo $\pi$ pulses.}
    \label{fig:Fig5}
\end{figure}

In Fig.~\ref{fig:Fig5}, we show the effect of multi-pulse PSE on NV~A, for which we calculate $M_{opt}=$ 3 and $\tau_{opt}=$ \SI{1.5}{\micro\second}. Here, $\tau$ is the interval between echo $\pi$ pulses, for a total free evolution time of $M\tau$. The additional pulses induce more complex time evolution, but also increase signal relative to the single-pulse PSE. The maximal $\Phi$ is approximately doubled on NV~A for the $M=3$ sequence relative to $M=1$. Because this system involves coherent interaction between probe and bath spins, this enhancement is distinct from applications where dynamical decoupling on a sensor qubit is used to increase sensitivity to a classical noise field. 

\begin{figure}
    \centering
    \includegraphics[width=80mm]{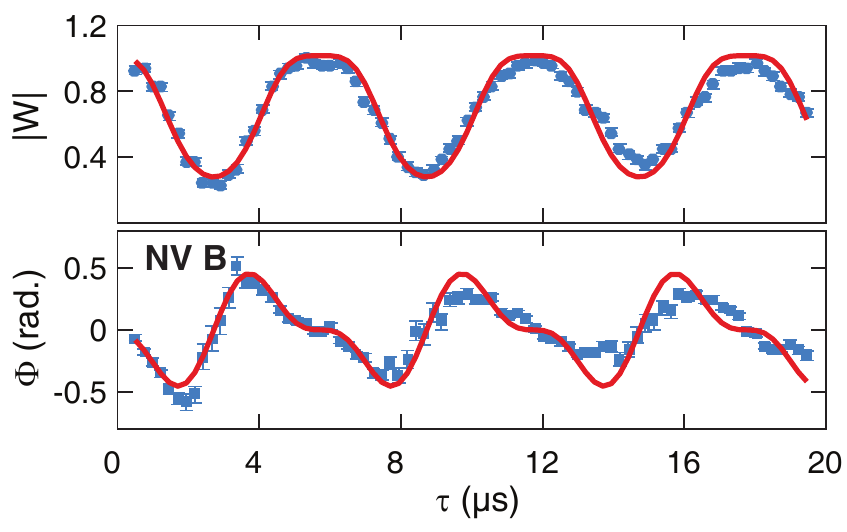}
    \caption{PSE measurement of the coherence magnitude $|W|$ and phase $\Phi$ after a single ($M=1$) echo with total evolution time $\tau$. When $\tau$ exceeds $|A_\parallel|^{-1}$, $(A_\perp)^{-1}$, the bath dynamics become increasingly non-Gaussian and the NV signal departs from predictions under the Gaussian approximation (red curves). As demonstrated here on NV~B, both $\langle X\rangle$ and $\langle Y\rangle$ begin to deviate, and PSE measurements can be used to quantify these non-Gaussian deviations.}
    \label{fig:Fig6}
\end{figure}

Finally, we note the potential utility of QPS measurements in exploring the underlying non-Gaussian nature of spin baths. While an ideal Gaussian environment would produce perfectly periodic oscillations, physical spin baths deviate from this ideal for $\tau > (A_\perp)^{-1},|A_\parallel|^{-1}$. These deviations are evident in both $|W|$ and $\Phi$, as observed in Fig.~\ref{fig:Fig6} using NV B, where the mismatch increases with $\tau$. By quantifying the non-Gaussian phase evolution, QPS measurements may enable tests of noise models for polarized and other non-equilibrium spin baths. Such investigations are outside the scope of the current work, but present a promising future direction of study.

\section{\label{sec:Discussion}Discussion}
We have used single NV centers in diamond to observe a periodic phase shift which arises in spin echo measurements due to axial polarization of surrounding nuclear bath spins. This quantum quench phase shift has been predicted previously through linear response calculations of Gaussian spin baths, and arises due to the bath Hamiltonian's dependence on the qubit state. We have extended the existing theory by calculating the effects of transverse polarization and multiple spin echo pulses, which are relevant to experimental implementations. A critical step in observing the quench phase was introducing a pulse sequence to minimize changes to the bath polarization due to many repetitions of the measurement protocol. 

QPS-based polarization measurements have many appealing characteristics. Not only do QPS measurements directly probe bath polarization $\bar{p}_z \in [-1,1]$, they can be performed with no prior knowledge of bath coupling parameters. This is possible since the empirically determined $\epsilon$ characterizes the total bath-qubit coupling, regardless of the number of spins or their distribution, so long as the environment is approximately Gaussian. In existing NV-based polarization measurement techniques, $^{13}$C bath polarization is probed indirectly via polarization loss from the central spin~\cite{Scheuer_PhysRevB2017}, is measured by addressing each nuclear spin through unique hyperfine couplings~\cite{cramer2016repeated}, or is detected by driving resonant nuclear spin rotations~\cite{Bucher_PhysRevX2020,Glenn_Nature2018}. Indirect measurements have difficulty distinguishing $p_z$ and $p_\perp$, while methods that drive nuclear spins entail quantum operations that last $\gg T_L$. In contrast, QPS measurements require only a single spin echo on the timescale of $T_L$, encompass the collective bath polarization in a single measurement, and do not require identifying or driving bath resonances. The Supplementary Materials contain additional data on existing polarization measurement techniques~\cite{supplement}.

Though not required in this work, the capability of tuning $\epsilon$ using the magnetic field is valuable for optimizing future spin bath measurements. For Gaussian spin baths with large $\epsilon$, such as systems with dense or strongly-coupled spins, $\epsilon$ can be moderated by increasing $B_0$. Conversely, lower applied fields may be use to probe spin baths at greater distances. Where approximate coupling strengths and spin distributions are known, Eq.~\eqref{eq:epsilondef} guides the choice of system parameters. As an example, to optimally probe the polarization of proton spins in a hydrocarbon liquid on the diamond surface using proximal NV centers roughly 7~nm deep, such as in Ref.~\cite{staudacher2013nuclear}, $B_0\approx$ 20 G can be used. Control over $\epsilon$ complements the ability to increase signal via multiple echoes; if $\epsilon$ is still small after tuning the magnetic field, multi-pulse measurements provide additional enhancement. Thus, QPS sequences are viable for probing a wide range of spin baths, with respect to both geometry and composition.

Engineering the spin sensor itself may also extend the capabilities of QPS measurements. Often, ensembles are used to increase signal-to-noise of single spin sensors. The simplest approach, using a randomly distributed ensemble to probe a randomly distributed spin bath, may not enhance QPS signal, due to averaging over the much weaker hyperfine coupling of more distant bath spins and the non-Gaussian behavior of proximal spins. However, more structured ensembles may yield improvements, such as for a thin layer of qubits near an interface to a target sample~\cite{devience2015nanoscale,vaidya2023hbn}. The similar distance of each sensor to the bath would produce a narrow distribution of $\epsilon$, ensuring that the ensemble measurements average to a consistent signal.

Simple, accurate methods for quantifying polarization are valuable in a variety of fundamental and applied domains. NV centers are actively used for developing efficient dynamic nuclear polarization methods~\cite{Schwartz_SciAdv2018}, evaluating hyperpolarization routines for quantum sensing~\cite{tetienne2021prospects}, and preparing registers of nuclear spins for quantum memories or quantum many-body experiments~\cite{takou2023precise}. Performing QPS measurements on near-surface NVs with a spin-rich target on the surface~\cite{staudacher2013nuclear}, in isotopically engineered diamond with layered architectures~\cite{Ohno_APL2012}, or with core-shell nanoparticles~\cite{Mindarava_JPhysChem2021} will provide more insight into promising sensing applications and the empirical limits of sensitivity for NV centers. Beyond NV centers, a rich variety of spin bath systems feature similar dynamics~\cite{witzel2007decoherence,seo2016quantum,murzakhanov2022electron}.

More broadly, QPS measurements provide an additional technique to study response properties of quantum systems, which is known to be a valuable probe of a variety of physical phenomena including superconductivity~\cite{Demler2022a,Demler2022b} and phase transitions~\cite{Machado2022}.
As discussed in~\cite{Wang_NatComm2021}, it can be used to quantitatively investigate bath dynamics and non-equilibrium systems, as well as  non-Gaussian noise. In a different context, this line of study may also shed light on the understanding of quantum-to-classical transitions~\cite{Kwiatkowski_NJOP2021,lin2022precession}.

\section*{\label{sec:Acknowledgements}Acknowledgements}
We thank Masaya Fukami, Jonathan C. Marcks, Leah Weiss, and Nazar Delegan for helpful discussions. This work is primarily supported by the U.S. Department of Energy, Office of Science, National Quantum Information Science Research Centers (Q-NEXT), with additional support from the U.S. Department of Energy, Office of Basic Energy Sciences, Materials Science and Engineering Division and the Center for Novel Pathways to Quantum Coherence in Materials, an Energy Frontier Research Center funded by the U.S. Department of Energy, Office of Science, Basic Energy Sciences under Award No. DE-AC02-05CH11231. This work was partially supported by the University of Chicago Materials Research Science and Engineering Center, which is funded by the National Science Foundation under award number DMR-2011854. A. C. also acknowledges support from the Simons Foundation through a Simons Investigator award (Grant No.~669487).

\appendix
\appsection{\label{sec:AppADerivations}Derivation of NV phase shifts from Gaussian baths}
\subsection{Probe qubit evolution due to a (quenched) Gaussian bath}

\label{appsec:qbcoh.gaus.gen}

We consider a Ramsey-type experiment (including Hahn echo or general dynamical decoupling sequences) on a single probe qubit. Throughout our discussion, we focus on the case where the qubit is coupled to a pure-dephasing environment. In this subsection, we provide a detailed discussion on the qubit phase shift and coherence decay effects due to a generic Gaussian bath (see~\cite{Wang_NatComm2021} for generalization to non-Gaussian environments).  
We also assume effects due to finite qubit pulse-width are negligible, which holds true for the experiments discussed in the main text. As such, we can transform to the standard toggling frame defined by the qubit pulses, as well as the rotating frame with respect to the intrinsic qubit Hamiltonian 
$ \hat H _{\mathrm{qb}} =
\omega _{\mathrm{qb}} 
\hat \sigma _{z} /2 
\equiv (\omega 
_{\mathrm{qb}}  /2) 
( \left |  \uparrow \rangle 
\langle \uparrow  \right | 
-  \left |  \downarrow  \rangle 
\langle  \downarrow \right | ) $. The toggling-frame Hamiltonian for the qubit-bath system can thus be written as 
\begin{align} 
\hat{ H} _{ 0 } 
\left ( t \right ) 
= & \hat{ H} _{\mathrm{env}} 
+ \hat{ H} _{\mathrm{int}} 
\left ( t \right )
, \\
\label{appaeq:H.int.gen}
\hat{ H} _{\mathrm{int}} 
\left ( t \right )
= & \frac{1}{2} 
F \left ( t \right )
\hat \sigma _{z} 
\otimes \hat \xi 
. 
\end{align}
Here, $\hat{ H} _{\mathrm{env}} $ denotes the bath-only Hamiltonian, $\hat \xi $ is the bath operator coupled to the qubit, and $F \left ( t \right )$ is the switching function that encodes qubit pulse(s). For the simplest case of Ramsey experiments, we have $F \left ( t \right ) =1$ during the protocol $t \in (0, t_{f} ]$.  

At the beginning of the protocol and after the qubit initialization ($\pi/2$) pulse, the instantaneous qubit-environment state is given by 
\begin{align}
& \hat{\rho}_{\mathrm{tot}} ( t = 0 ^+ )
=  \left |  + \rangle 
\langle + \right | 
\otimes 
\hat{\rho}_{\mathrm{b,i}} 
	,  
\end{align}
where $\left |  + \right \rangle 
\equiv \frac{1}{\sqrt{2} } 
\left( 
\left |  \uparrow \right \rangle 
+\left |  \downarrow \right \rangle 
\right )$.
By measuring the qubit spin operators 
$\hat \sigma _{x} $ and $\hat \sigma _{y} $ at the end of the protocol $ t = t_f $, one can directly access the qubit coherence function, defined as follows ($\hat \sigma _- 
\equiv  \left |  \downarrow  \rangle 
\langle  \uparrow \right |
= (\hat \sigma _{x} 
- i \hat \sigma _{y} 
)/2 $)
\begin{align}
\label{appaeq:qb.coh.tf.gen}
& \langle 
	\hat \sigma _- ( t_f  ) 
	\rangle
\equiv \textrm{Tr} 
\left [ \left ( \hat \sigma _- 
\otimes \hat {\mathbb{I} }
_{\mathrm{env}}  
\right )
\hat{\rho}_{\mathrm{tot}} ( t_f )
\right ]
, \\
\label{appaeq:rho.tot.tf}
& \hat{\rho}_{\mathrm{tot}} ( t_f )
    = \mathcal{T} 
    e^{-  i \int_0^{t_f} dt' 
    \hat H_{ 0 }  (t') }
	\hat{\rho}_{\mathrm{tot}} ( t = 0 ^+ )
	\tilde{\mathcal{T}} 
    e^{   i \int_0^{t_f} dt' 
    \hat H_{ 0 }  (t') } 
	. 
\end{align}
Due to the pure-dephasing form of qubit-environment coupling in Eq.~\eqref{appaeq:H.int.gen}, the qubit coherence function in Eq.~\eqref{appaeq:qb.coh.tf.gen} can be computed via the initial bath state and conditional bath Hamiltonians 
$ \hat H _{ \uparrow (\downarrow) } (t)
    \equiv \hat{ H} _{\mathrm{env}} 
     \pm \frac{1}{2} F (t) \hat{\xi}$, as
\begin{align}
\label{appaeq:qb.coh.evo}
& \frac{\langle 
	\hat \sigma _- ( t_f  ) 
	\rangle}
	{\langle 
	\hat \sigma _- ( 0^+  ) 
	\rangle}
 = \text{Tr} \left\{
    \mathcal{T} 
    e^{-  i \int_0^{t_f} dt' 
    \hat H _{ \uparrow  }  (t')}
    \hat{\rho}_{\mathrm{b,i}} 
    \tilde{\mathcal{T}} 
    e^{   i \int_0^{t_f} dt' 
    \hat H _{ \downarrow  }  (t') }
    \right\} 
     . 
\end{align}
It is thus convenient to separate the dephasing from the phase shift effect in the measurement signal, so that we can define 
\begin{align}
\label{appaeq:qb.coh.log}
\frac{\langle 
	\hat \sigma _- ( t_f  ) 
	\rangle}
	{\langle 
	\hat \sigma _- ( 0^+  ) 
	\rangle}
	&  =  \,
	e^{- \chi (t_f)
 - i \Phi(t_f)}  
	.
\end{align}
In Eq.~\eqref{appaeq:qb.coh.log}, the real-valued functions $\chi (t_f)$ and $\Phi(t_f)$ encode amplitude decay and phase shift on the qubit coherence due to coupling to the environment, respectively.

As shown in Ref.~\cite{Wang_NatComm2021}, implementation of a standard dephasing-type noise spectroscopy experiment can inadvertently lead to a quench on the environment during the protocol, resulting in an extra quench-induced qubit phase shift. To see this effect, we can transform to the interaction picture defined by the bath Hamiltonian $\hat{H}_{\mathrm{b,i}} $ prior to the start of the measurement sequence. More specifically, for experiments discussed in the main text, we have 
\begin{align}
& \hat{H}_{\mathrm{b,i}} 
= \hat H _{ \uparrow } (t = 0^{-})
= \hat{ H} _{\mathrm{env}} 
+ \frac{1}{2} \hat{\xi}  
     .  
\end{align}
The corresponding interaction-picture Hamiltonian with respect to $\hat{H}_{\mathrm{b,i}} $ can be written as 
\begin{align} 
\hat{ H} _{\mathrm{I}} 
\left ( t \right )
= -\frac{1}{2} \hat{\xi} (t) 
+ \frac{1}{2} 
F \left ( t \right )
\hat \sigma _{z} 
\otimes \hat \xi 
\left ( t \right ) 
,  
\end{align}
where we define the rotating-frame bath operator $\hat{\xi} (t) $ as 
\begin{align} 
\hat{\xi} (t) \equiv 
e^{ i \hat{H}_{\mathrm{b,i}} t } 
\hat{\xi}  
e^{- i \hat{H}_{\mathrm{b,i}} t } 
.  
\end{align}
Thus Eq.~\eqref{appaeq:qb.coh.evo} can be equivalently computed in the interaction picture, as 
\begin{align}
\label{appaeq:qb.coh.evo.int}
\frac{\langle 
	\hat \sigma _- ( t_f  ) 
	\rangle}
	{\langle 
	\hat \sigma _- ( 0^+  ) 
	\rangle}
& = \text{Tr} \left\{
    \hat U _{\mathrm{I}, \uparrow  } 
\left ( t_f \right )
    \hat{\rho}_{\mathrm{b,i}} 
    \hat U _{\mathrm{I},\downarrow  } 
    ^{\dag } 
\left ( t_f \right )
    \right\} 
    ,
\\
\hat U _{\mathrm{I},\uparrow / \downarrow } 
\left ( t_f \right )
& \equiv 
\mathcal{T} 
    e^{ \pm \frac{i}{2} 
    \int_0^{t_f} dt' 
    \left [ F \left ( t \right )
    \pm 1 \right ]
    \hat \xi \left ( t' \right ) 
    }
\end{align}
If the environment is Gaussian, i.e.~if the bath operator $\hat \xi 
\left ( t \right ) $ satisfies Gaussian statistics,  Eq.~\eqref{appaeq:qb.coh.evo.int} can be evaluated exactly. For this purpose, we first note that a generic Gaussian process can be fully characterized by its first-order average and second-order correlation functions.  For the bath noise operators $\hat \xi 
\left ( t \right ) $, we thus introduce   
\begin{align} 
\langle 
\hat{\xi}  (t  ) 
\rangle & \equiv  
\text{Tr}  [
    \hat{\xi} (t) 
    \hat{\rho}_{\mathrm{b,i}} 
    ] 
, \\
\langle \delta 
\hat{\xi} ( t _{1})
\delta \hat{\xi} ( t _{2} )
\rangle
&  \equiv  
\text{Tr} [
    \delta  \hat{\xi} (t _{1}) 
    \delta  \hat{\xi} (t _{2}) 
    \hat{\rho}_{\mathrm{b,i}} 
]
,   
\end{align} 
where $\delta  \hat{\xi} (t) 
\equiv \hat{\xi} (t) 
- \langle 
\hat{\xi}  (t  ) 
\rangle 
$.
The first order average 
$\langle \hat{\xi}  (t  ) 
\rangle $ can be viewed as the quantum version of the average of a classical stochastic field. 
The second-order average 
$\langle \hat{\xi} ( t _{1} )
\hat{\xi} ( t _{2} )
\rangle$, in contrast to the case of classical stochastic variables, is generally not symmetric under exchange of time variables $t _{1}$ and $t _{2}$, because the bath operators $\hat{\xi} ( t )$ at different times do not commute. In this case, a more physical way to represent the second-order moments is to separate its symmetric- and asymmetric-in-time components. Specifically, we can define the standard bath autocorrelation function 
$\bar{S} (t _{1},t _{2})$, as well as its linear response susceptibility function $G ^R _{\xi \xi } (t _{1},t _{2}) $, as  
\begin{align} 
\bar{S} (t _{1},t _{2})
	 & \equiv \frac{1}{2}  
 \langle
	\{ \delta  \hat{\xi} ( t _{1} )
	,
	\delta  \hat{\xi} ( t _{2} )
	\}
	\rangle
,\\
G ^R _{\xi \xi } (t _{1},t _{2})
    & \equiv 
    \! 
    -i \Theta (t _{1}-t _{2})
    \langle
    [\hat{\xi} ( t _{1} )
	    ,
    \hat{\xi} ( t _{2} )]
    \rangle
     . 
\end{align}

Making use of the first two bath average moments, we can now explicitly write out the phase shift $\Phi(t_f)$ and dephasing factor $\chi (t_f)$ in Eq.~\eqref{appaeq:qb.coh.evo} due to a Gaussian environment. The dephasing term is given by  
\begin{align} 
\label{appaeq:qb.deph.gaus}
\chi (t_f)	&  = 
\!  \int_0^{ t_f  } \!   d t_1 F(t_1) 
    \! \int_0^{ t_1  }  d t_2 F(t_2)    \bar{S} (t _{1},t _{2})
\end{align}
As such, the dynamical decoupling pulses encoded in $ F (t)$ act as a spectral filter of the bath noise, and the corresponding $F[\omega]$ in the Fourier domain (or its squared norm) is commonly known as the filter function of the pulse(s). Note that Eq.~\eqref{appaeq:qb.deph.gaus} has a direct parallel with qubit dephasing due to classical noise. For the phase shift $\Phi(t_f)$, we have  
\begin{align} 
\label{appaeq:qb.phase.gaus}
 \Phi(t_f) & = 
\Phi _{m}(t_f) 
+ \Phi _{q}(t_f) 
, \\
\label{appaeq:qb.ph.gaus.lin}
\Phi _{m}(t_f) 
& = 
\int_0^{ t_f  } \!   d t_1 F(t_1) \langle \hat{\xi}  (t _1 ) 
\rangle 
, \\
\label{appaeq:qb.gaus.qps}
\Phi _{q}(t_f) & = 
- \frac{1}{2}
\!  \int_0^{ t_f  } \!   d t_1 F(t_1) 
    \! \int_0^{ t_1  }   d t_2   
    G ^R _{\xi \xi  } (t_1 , t_2)
    .
\end{align}
In the RHS of Eq.~\eqref{appaeq:qb.phase.gaus}, the first term $\Phi _{m}(t_f) $ corresponds to the standard phase shift due to a nonzero average bath field, which is independent of the probe qubit state and also has a straightforward classical analog. The second quench-induced term $\Phi _{q}(t_f) $, however, is related to the response properties of the quantum environment, and cannot be generated by a classical static noise source. This extra phase shift, which we call the quench phase shift (QPS) for convenience, can emerge in Ramsey-type experiments for a wide range of physical platforms, which in turn can be used to extract the spectral function of an unknown environment~\cite{Wang_NatComm2021}.


\subsection{NV dynamics under multipulse dynamical decoupling sequences}

\label{appsec:qbcoh.13C.CPMG}

We note that Eqs.~\eqref{appaeq:qb.deph.gaus} and~\eqref{appaeq:qb.gaus.qps} are generally applicable to computing the dephasing factor $ \chi (t_f) $ and the quench phase shift $ \Phi _{q} (t_f) $ for general dynamical decoupling pulses. As an example, we derive NV dynamics due to a nuclear spin environment corresponding to a general Carr-Purcell-Meiboom-Gill (CPMG) pulse sequence with $M$ pulses and total duration $M \tau$, as considered in the main text. The pulse sequence can now be written as $\left[ \frac{\tau}{2} - \pi - \frac{\tau}{2}\right]^M$, and the switching function can be compactly expressed as
\begin{align}
\label{appaeq:Ft.CPMG}
F (t) = \text{sgn} 
\left[ 
\cos\left(
\pi  t / \tau 
\right) \right]
, \quad
0 \le t \le M \tau 
, 
\end{align}
where $\text{sgn} (\cdot)$ denotes the sign function.

We now consider nuclear spin bath discussed in the main text. As shown in the supplement, the bath Hamiltonian and NV-bath coupling operator can be described by
\begin{align} 
\label{appaeq:13c.h.bi}
\hat{H}_{ \mathrm{b,i} } 
& = \omega _{L} 
\sum _{ j }
\hat I _{z,  j } 
,   \\
\label{appaeq:13c.hxi}
\hat \xi & = 
- \sum _{ j }
\left ( A _{\parallel, j }
\hat I _{z, j }
+ A _{\perp,  j }
\hat I _{x,  j  }
\right ) 
. 
\end{align}
Without loss of generality, we also assume the nuclear spins only have axial polarization. In this case, by substituting Eq.~\eqref{appaeq:Ft.CPMG} into the analytical expressions for dephasing and phase evolution in Eqs.~\eqref{appaeq:qb.deph.gaus} and~\eqref{appaeq:qb.gaus.qps}, we obtain 
\begin{align}
\chi (\tau  ) & = 
\frac{ \sum _{ j }  A _{\perp,  j } ^{2 }  }
{ \omega _{L} ^{2}} 
\frac{\sin^4 \frac{\omega_{L} \tau}{4 } }
{2 \cos^2 \frac{\omega_L \tau}{2 }}
\left | (-) ^{M+1}
e ^{i \omega_L M \tau}
+1 
\right | ^{2}
, \\
\Phi_{q}(\tau  ) & =(-1)^{M-1}
\frac{ \sum _{ j } p _{z,  j  } 
A _{\perp,  j } ^{2 } } 
{ \omega  _{L} ^{2}}  
\frac{\sin{M\omega_{L}\tau}\sin^2
\frac{\omega_{L}\tau}{4}}
{2\cos\frac{\omega_{L}\tau}{2}}
. 
\end{align}
One can show that above equations are equivalent to Eqs.~\eqref{eq:multipulseChi} and~\eqref{eq:multipulsePhi} in the main text. 

\subsection{Derivation of the upper bound on NV $\langle \hat \sigma_y \rangle$ coherence signal due to QPS}

\label{appsec:qbcoh.13C.sy.lim}

Let us again consider nuclear spin bath discussed in the main text, but with a generic dynamical decoupling pulse satisfying 
$ \int _{ 0 }^{ t_{f} } 
\!  F (t) dt =0 $ (i.e.,~any static noise is fully canceled). Substituting Eqs.~\eqref{appaeq:13c.h.bi} and \eqref{appaeq:13c.hxi} into the general analytical expressions for dephasing and phase evolution in Eqs.~\eqref{appaeq:qb.deph.gaus} and~\eqref{appaeq:qb.gaus.qps}, and introducing the Fourier transform of filter function as
\begin{align}
F[\omega]  &  \equiv 
\int _{ 0 }^{ t_{f} } 
\!  F (t) 
  e^{ i \omega t} d t 
, 
\end{align}
we thus obtain
\begin{align}
\label{appaeq:qb.deph.1freq}
\chi ( t_f  ) & = 
\frac{ \sum _{ j }  A _{\perp,  j } ^{2 }  }{ 8} 
\left |F[\omega _{L} ] 
\right| ^{2}
, \\
\label{appaeq:qb.qps.1freq}
\Phi_{q}( t_f ) & = 
\frac{ \sum _{ j } p _{z,  j  } 
A _{\perp,  j } ^{2 } } 
{ 4 \omega  _{L} }  
\text{Re}F[\omega _{L} ] 
.
\end{align}
One can use a few lines of algebra to show that Eqs.~\eqref{appaeq:qb.deph.1freq} and~\eqref{appaeq:qb.qps.1freq} lead to an upper bound on the NV coherence component $\langle \hat \sigma_y \rangle$
\begin{align}
| \langle \hat \sigma_y \rangle| 
=& e^{-\chi( t_f ) }
|\sin \Phi_{q}( t_f ) | 
\le |\Phi_{q}( t_f )|
e^{-\chi( t_f ) } 
\nonumber 
\\
\le & \frac{1}{2 \sqrt{e} }
\sqrt{\frac{ \sum _{ j }  A _{\perp,  j } ^{2 } }
{ \omega _{L} ^{2 }} }  
, 
\end{align}
which reproduces the upper bound given in the main text.

\appsection{\label{sec:AppBTransPol}Origins of transverse polarization}
In standard dynamical nuclear polarization protocols and in the simplest scenarios, which typically make use of Hartmann-Hahn resonances, it is common to assume that the resulting nuclear polarization is aligned with the external magnetic field, or equivalently, the direction of its bare Hamiltonian. It is thus useful to discuss the origin of both parallel and transverse polarizations in the experiments discussed in the main text. Let us again start with the total system Hamiltonian, which can be written as
\begin{align}
\label{appbeq:nv.13c.htot}
& \hat{H}_{\mathrm{tot}}  
=  \frac{\omega _{\mathrm{NV}} }
{2}
\hat \sigma _{z} 
+ \hat{H}_{ \mathrm{b,i} } 
+  \frac{\hat {\mathbb{I} } 
_{\mathrm{NV}} + \hat \sigma _{z}}{2}
\otimes 
( \hat{H}_{ \mathrm{b,i} }  
+ \hat \xi  ) 
. 
\end{align}
For clarity, in the following derivations we assume the NV is coupled to a single nuclear bath spin (with spin operators given by $\hat \sigma  _{0, \alpha } $), but our result also applies to larger spin baths. Further, we also show numerical evidence verifying our analytical results, assuming a realistic spin bath corresponding to NV~A discussed in the main text.

To transfer polarization from the NV to the nuclear spin, we make use of a spin locking pulse on the NV. Without loss of generality, we assume the drive is along the $x$ axis, which can be described in the NV Hamiltonian as 
$\hat{H}_{\mathrm{dr}} 
= \Omega _{\mathrm{dr}} \hat \sigma _{x} 
\cos (\omega _{\mathrm{dr}} t )$, with the drive frequency and  Rabi amplitude given by 
$\omega _{\mathrm{dr}} $ and 
$\Omega _{\mathrm{dr}}$ respectively. In the rotating frame defined with respect to $\omega _{\mathrm{dr}} $, the total NV-bath system can be described by the following Hamiltonian 
\begin{align}
\label{appeq:Hamiltonian.SL.0}
\hat{H}_{\mathrm{SL}} 
& = - \frac{\delta}{2}
\hat \sigma _{z} 
+ \frac{\Omega _{\mathrm{dr}} }{2}
\hat \sigma _{x} 
+ \hat{ H} _{\mathrm{bath}} 
+ \hat{ H} _{\mathrm{int}} 
, \\
\hat{ H} _{\mathrm{bath}}  
& = \frac{\omega _{L}  }{2}
\hat \sigma _{z,0}
, \\
\hat{ H} _{\mathrm{int}} 
& = \frac{ \hat {\mathbb{I} } 
_{\mathrm{NV}} 
- \hat \sigma _{z} }{2}
\otimes 
\frac{ A _{\parallel, 0}
\hat \sigma _{z,0} 
+  A _{\perp, 0}
\hat \sigma _{x,0} }{2}
, 
\end{align}
where the drive detuning is defined as 
$\delta \equiv 
\omega _{\mathrm{dr}}  - 
\omega _{\mathrm{NV}} $. To enable polarization transfer from NV spin to the bath spin, we choose the spin locking pulse to satisfy resonance conditions $\delta = 0$ and 
$\Omega _{\mathrm{dr}}  
= \omega _{L}  $, so that the NV-bath Hamiltonian simplifies into 
\begin{align}
\label{appbeq:novel.htot}
& \hat{H}_{\mathrm{SL}} 
= \frac{\omega _{L} }{2}
\left( \hat \sigma _{x} 
+ \hat \sigma _{0, z} 
\right )
+ \hat{H}_{\mathrm{quench}} 
+ \hat{H}_{\mathrm{int}} 
, \\
\label{appbeq:h.sl.quench}
& \hat{H}_{\mathrm{quench}} 
= \frac{ 1 }{4} 
\left( A _{\parallel, 0}
\hat \sigma _{0, z} 
+  A _{\perp, 0}
\hat \sigma _{0, x}
\right )
, \\
\label{appbeq:h.sl.int}
& \hat{H}_{\mathrm{int}} 
= - \frac{ 1 }{4} 
\hat \sigma _{z}  
\otimes 
\left( A _{\parallel, 0}
\hat \sigma _{0, z} 
+  A _{\perp, 0}
\hat \sigma _{0, x}
\right )
. 
\end{align}

\begin{figure}
    \centering
    \includegraphics[width=83mm]{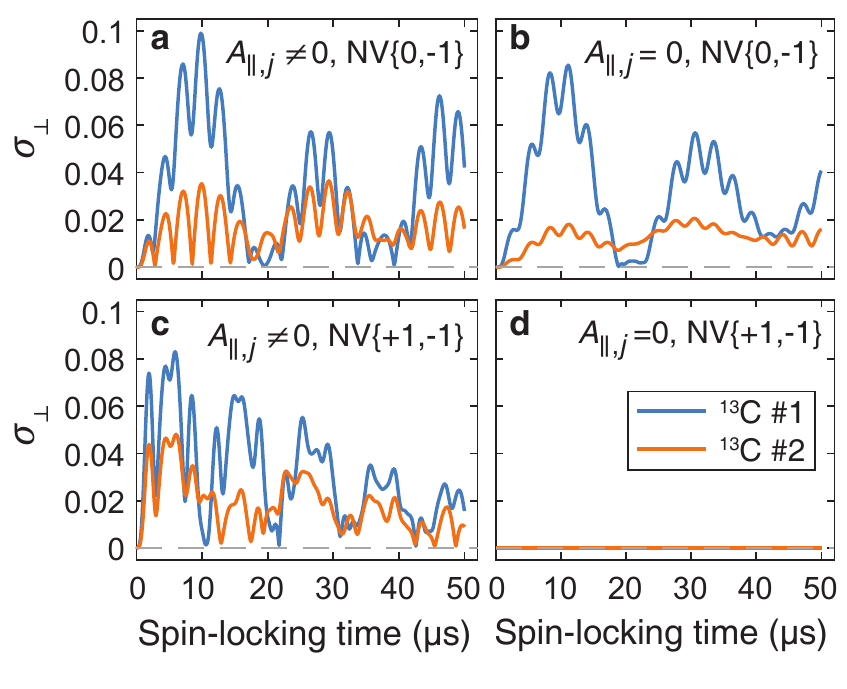}
    \caption{Origin of transverse nuclear spin polarization. Plotted is the time evolution of $^{13}$C \#1 and \#2, when simulating the entire NV-nuclear spin bath system dynamics, making use of fitted hyperfine coupling parameters of NV~A from Table~\ref{tab:Table1}. Panel \textbf{a} shows simulation results based on the spin locking Hamiltonian Eq.~\eqref{appeq:Hamiltonian.SL.0}, whereas the panels \textbf{b}-\textbf{d} depict simulations when ignoring the axial component of hyperfine couplings, when using a balanced NV $|m _{s}=+1 \rangle \leftrightarrow |m _{s}=-1 \rangle $ transition in the spin locking pulse, and when applying both. While only two $^{13}$C spins are plotted here, the polarization dynamics of the other $^{13}$C bath spins exhibit qualitatively the same behavior. }
    \label{appfig:nuc.perp.pol}
\end{figure}

We note that when the nuclear spin Larmor frequency $\omega _{L} $ is comparable to 
$A _{\perp, 0} $ or $A _{\parallel, 0} $, the spin-locking Hamiltonian $\hat{H}_{\mathrm{SL}} $ may not facilitate perfect polarization transfer, i.e.~$\hat{H}_{\mathrm{SL}} $ in general cannot transform an unpolarized initial bath spin state to $\hat \sigma _{0, z} $ eigenstates. This can be seen intuitively by comparing the physical Hamiltonian $\hat{H}_{\mathrm{SL}} $ to the standard flip-flop Hamiltonian for $2$-spin system, the latter of which can achieve perfect polarization transfer, as  
\begin{align}
\label{appbeq:h.flip}
& \hat{H}_{\mathrm{swap}} 
= \frac{\omega _{L} }{2}
\left( \hat \tau _{1,x} 
+ \hat \tau _{2, z} 
\right )
+ J \left( \left |  -  \rangle _{1}
\langle +  \right | 
\otimes 
\hat \tau _{2, +} 
+ \text{H.c.} 
\right )
. 
\end{align}
We see that there are two different factors that prevent Eq.~\eqref{appbeq:novel.htot} from achieving perfect polarization transfer: 
\begin{enumerate}
    \item{The NV-nuclear spin interaction $\hat{H}_{\mathrm{int}} $ deviates from the perfect flip-flop form by having the extra $A _{\parallel, 0} $ term as well as the counter-rotating contributions in the $A _{\perp, 0} $ term. } 
    \item{The extra quench term $\hat{H}_{\mathrm{quench}} $ tilts the nuclear spin axis, so that the intrinsic nuclear spin quantization axis now has overlap with the direction set by the interaction Hamiltonian $\hat{H}_{\mathrm{int}} $ between the NV and the bath. } 
\end{enumerate}
This analysis is also verified by numerical simulations in Fig.~\ref{appfig:nuc.perp.pol}, where we assume approximately the experimental magnetic field $B_0 = 312~\text{G}$, and the nuclear spin configuration of NV~A with $5$ bath spins extracted from XY8-$2$ measurements (see Table~1 in the main text). We start from an unpolarized initial nuclear spin state, and simulate the time evolution of nuclear spin polarization along $z$ and $x$ axes during a single spin-locking pulse (Fig.~\ref{appfig:nuc.perp.pol}a). To demonstrate the two distinct sources of transverse polarization discussed above, we also simulate the spin polarization evolution for the same bath but driving NV $|m _{s}=+1 \rangle \leftrightarrow |m _{s}=-1 \rangle $ transition in the spin locking pulse (Fig.~\ref{appfig:nuc.perp.pol}c), in which case the total Hamiltonian does not have the quench term in Eq.~\eqref{appbeq:h.sl.quench}. We compare these results with (i) setting the parallel hyperfine coupling coefficients to zero $A _{\parallel, 0} \equiv 0$ in Eq.~\eqref{appbeq:h.sl.int}, and more generally $A _{\parallel, j} \equiv 0$ for the multispin bath used in the simulation (Fig.~\ref{appfig:nuc.perp.pol}b), such that the nuclear spin quantization axis is not tilted from the $z$ axis, versus (ii) using both NV $|m _{s}=+1 \rangle \leftrightarrow |m _{s}=-1 \rangle $ transition and the transverse-only NV-bath interaction (Fig.~\ref{appfig:nuc.perp.pol}d). As shown in Fig.~\ref{appfig:nuc.perp.pol}, all but the last case lead to nontrivial transverse polarization in the $xy$ plane, illustrating that having either the quench term Eq.~\eqref{appbeq:h.sl.quench} or the counter-rotating contributions in Eq.~\eqref{appbeq:h.sl.int} can prevent the bath from reaching fully polarized state during the spin-locking pulse. 

It is worth noting that in the limit where rotating wave approximation holds, 
i.e.~$\omega _{L} 
\gg |A _{\parallel, 0} |, |A _{\perp, 0} | $, $\hat{H}_{\mathrm{SL}} $ in Eq.~\eqref{appbeq:novel.htot} can be well approximated by Eq.~\eqref{appbeq:h.flip}, and the dynamics could achieve perfect polarization transfer. However, in this regime the polarization transfer process also becomes much slower than the timescale set by nuclear spin Larmor frequency, which can significantly prolong the entire measurement protocol. In realistic experiments, it would be more desirable to use nuclear spins with larger coupling, i.e.~greater values of $|A _{\perp, 0} |$ to ensure that the later sensing step can detect reasonable signal.

\appsection{Analysis of the compensating spin echo sequence}
\label{sec:AppC_CSE}
In the main text, we state that the phase-resolved spin echo signal is more robust, if we apply a second, compensating spin echo pulse using NV basis $\{m _{s}=0,m _{s}=+1\}$, and fix the distance between the start of the two Hahn echo sequences to be integer multiples of the nuclear Larmor period 
$T_L=2\pi \omega_L^{-1}$, as illustrated in Fig.~\ref{fig:Fig3}a. Here, we provide a rigorous justification for the use of this measurement protocol.

Before explaining how the modified protocol offers a more robust approach to measuring the quench phase shift $ \Phi _{q} $ and the linear-order phase shift $ \Phi _{m} $, we first discuss why the standard spin echo measurement may be insufficient for this purpose. Note that in the ideal case considered in Ref.~\cite{Wang_NatComm2021}, where at the beginning of the measurement protocol, the sensing target relaxes into a steady state set by its surrounding environment, one can simply vary the wait time $t _{wait}$ between the end of the initialization pulse and the start of the spin echo sequence to access NV phase shifts with same axial but varying transverse polarizations (see Supplementary Materials for detail). For the experiment considered here, however, because we use the interaction between the probe NV and the nuclear spin bath to also initialize (i.e.~polarize) the bath state, the resulting initial bath state could also depend on 
$t _{wait}$ (when repeating the entire measurement cycle many times), making it difficult to separate the phase shifts due to axial and transverse polarizations. 

Thus, our goal is to devise a measurement protocol where the bath initialization is independent of the measurement pulse sequence, and more specifically, insensitive to $t_{{wait}}$. Fortunately, this can be achieved by the application of the second compensating Hahn echo pulse. As shown in Fig.~\ref{fig:Fig3} in the main text, the evolution of spin echo phase without the compensating sequence averaged over all $t_{ {wait}}$ (Fig.~\ref{fig:Fig3}c) shows considerable deviation from the analytical expression (see Supplementary Materials for detailed derivations)
\begin{align}  
&\left . 
\Phi _{q}(t_f)  
\right | 
_{ p _{x,  j  } = 
p _{y,  j  } = 0 }
\nonumber\\
\label{seq:qps.z}
= & \sum _{ j } p _{z,  j  } 
\frac{ A _{\perp,  j } ^{2 }   }
{ \omega  _{L}  ^{2 }} 
\sin  ^{2 } 
\frac{ \omega  _{L} t _{f}   }{4} 
\sin  \frac{ \omega  _{L} t _{f}   }{2} 
, 
\end{align}
which is due to drifting of initial bath polarization when we change the wait time $t _{wait}$. In contrast, adding the compensating pulse leads to excellent agreement between measured spin echo phase averaged over $t _{wait}$ (see solid curves in Fig.~3(e)) and the analytical expression in Eq.~\eqref{seq:qps.z}.  

One can naturally ask if there exists a more rigorous proof showing why the sequence with the compensating pulse is more robust, beyond numerics based on the specific spin bath configurations. To see this, it is convenient to transform the bath to the rotating frame defined with respect to the bath-only Hamiltonian, 
$ \hat{H}_{ \mathrm{b,i} } 
= \omega _{L} 
\sum _{j}
\hat I _{z, j}  $, as well as the toggling frame defined by the Hahn echo $\pi$ pulse. Thus, the action of a single Hahn echo sequence (with wait time $t$ and duration $\tau$) on the bath density matrix can be written as  
\begin{align}
& \mathcal{C} _{\mathrm{mhe} } 
\left ( t  ;  \tau \right )
    \left [ 
\hat{\rho}_{\mathrm{b}} \right ] 
\nonumber \\
= & 
\text{Tr} _{\mathrm{NV}} 
\left \{ \hat U 
_{ \{ 0,-1\} }
\left ( t  ;  \tau \right ) 
( \left | + \rangle 
\langle +  \right | 
\otimes \hat{\rho}_{\mathrm{b}} ) 
\hat U _{ \{ 0,-1\} }^{\dag}
\left ( t  ;  \tau \right ) 
 \right \} 
, 
\end{align}
where $\hat U _{ \{ 0,-1\} }$ denotes the unitary evolution operator of the total NV-bath system when using NV $\{m _{s}=0,m _{s}=-1\}$ basis as the probe qubit, and $\left | + \right \rangle $ is the equal superposition state of the two basis states. For the purpose of our discussion, it is convenient to also define the unitary evolution when using NV basis states $\{m _{s}=0,m _{s}=+1\}$ as $\hat U _{ \{ 0,+1\} }$, so that we can compactly derive the unitary operators as
\begin{align}
& \hat U _{ \{ 0,\pm 1\}  }
\left ( t  ;  \tau \right ) 
\nonumber \\
\equiv &  \mathcal{T} 
    e^{ \pm i \int_{ t + \frac{\tau}{2} }
    ^{t + \tau }  
    \left |  0 \rangle 
\langle 0 \right | 
\otimes 
\hat \xi  
\left ( t' \right ) dt'  }
\mathcal{T} 
    e^{ \pm i \int_{ t }
    ^{t + \frac{\tau}{2} }  
    \left |  \pm 1 \rangle 
\langle \pm 1 \right | 
\otimes 
\hat \xi  
\left ( t' \right ) dt'  }
.
\end{align}
Here, 
$\hat \xi  
\left ( t \right ) 
\equiv e ^{ i \hat{H}_{ \mathrm{b,i} } t} 
\hat \xi 
e ^{- i \hat{H}_{ \mathrm{b,i} } t} $. Note that 
$\mathcal{C} _{\mathrm{mhe} } 
\left ( t  ;  \tau \right ) $ denotes a superoperator, which is in general a completely positive trace preserving (CPTP) map. 

From the NV-bath system described by Eqs.~\eqref{appaeq:13c.h.bi} and~\eqref{appaeq:13c.hxi}, we obtain
$ \hat \xi (t) = 
- \sum _{ j } [ A _{\parallel,j }
\hat I _{z,j}
+ A _{\perp, j}
(\hat I _{+, j } e ^{i \omega _{L} t}
+  h.c. ) ] $; assuming the evolution time is sufficiently small such that 
$|A_{\parallel,j}| \tau  \ll 1 $ and $ |A_{\perp,j}| \tau \ll 1 $, we can approximately express the backaction due to the Hahn echo sequence on the nuclear spin bath as (keeping leading-order terms in 
$\hat \xi  \left ( t \right ) $)
\begin{align}
& \mathcal{C} _{\mathrm{mhe} } 
\left ( t  ;  \tau \right )
    \left [ 
\hat{\rho}_{\mathrm{b}} \right ] 
\simeq \hat{\rho}_{\mathrm{b}} 
- \frac{i }{2}
\int_{t  } ^{t + \tau}
    [\hat \xi  \left ( t' \right ) 
    , \hat{\rho}_{\mathrm{b}} ]
    dt'  
\nonumber \\
\label{appceq:mhe.approx}
\simeq  & 
\mathcal{T} 
    e^{ - \frac{i }{2}
    \int_{t  }
    ^{t + \tau} dt'  
    \hat \xi  
    \left ( t' \right ) 
    }
    \hat{\rho}_{\mathrm{b}} 
    \tilde{\mathcal{T}} 
    e^{ + \frac{i}{2} \int_{t  }
    ^{t + \tau } dt'  
    \hat \xi  
    \left ( t' \right )  } 
    . 
\end{align}
It is clear that the evolution of bath state under the Hahn echo measurement sequence depends on the value of $t $, i.e.~the wait time $t _{wait}$. 

This formal argument also lets us understand the role of the compensating Hahn echo pulse. Noting that the second Hahn echo sequence uses NV $\{m _{s}=0,m _{s}=+1\}$ basis, the evolution of bath state under the combination of measurement and compensating Hahn echo sequences can be written as  
\begin{align}
\label{appceq:2he.def}
& \mathcal{C} _{\mathrm{2he} } 
\left ( t _{1}, t _{2}  
;  \tau \right )
\left [ 
\hat{\rho}_{\mathrm{b}} \right ] 
= \mathcal{C} _{\mathrm{che} } 
\left ( t _{1} + t _{2}
;  \tau \right )
    \left [ 
\mathcal{C} _{\mathrm{mhe} } 
\left ( t _{ 1 } ;  
\tau \right )
    \left [ 
\hat{\rho}_{\mathrm{b}} \right ]  \right ] 
, 
\end{align}
where we have 
\begin{align}
& \mathcal{C} _{\mathrm{che} } 
\left ( t  ;  \tau \right )
    \left [ 
\hat{\rho}_{\mathrm{b}} \right ] 
\nonumber \\
= & \text{Tr} _{\mathrm{NV}} 
\left \{ \hat U 
_{ \{ 0,+ 1\}  }
\left ( t  ;  \tau \right ) 
( \left | + \rangle 
\langle +  \right | 
\otimes \hat{\rho}_{\mathrm{b}} ) 
\hat U _{ \{ 0,+ 1\}  }^{\dag}
\left ( t  ;  \tau \right ) 
 \right \} 
. 
\end{align}
In Eq.~\eqref{appceq:2he.def}, $t_{1}$ denotes the start of the first measurement Hahn echo pulse, whereas $t_{2}$ denotes the spacing between the start of the two Hahn echo sequences. Following similar analysis as in Eq.~\eqref{appceq:mhe.approx}, we can approximate the backaction due to the compensating pulse as 
\begin{align}
\label{appceq:che.approx}
\mathcal{C} _{\mathrm{che} } 
\left ( t  ;  \tau \right )
    \left [ 
\hat{\rho}_{\mathrm{b}} \right ] 
\simeq \mathcal{T} 
    e^{ + \frac{i }{2}
    \int_{t  }
    ^{t + \tau} dt'  
    \hat \xi 
    \left ( t' \right ) 
    }
    \hat{\rho}_{\mathrm{b}} 
    \tilde{\mathcal{T}} 
    e^{ - \frac{i}{2} \int_{t  }
    ^{t + \tau } dt'  
    \hat \xi  
    \left ( t' \right )  } 
    . 
\end{align}
Combining Eq.~\eqref{appceq:mhe.approx} and~\eqref{appceq:che.approx}, it is straightforward to see that the leading order effects due to the two Hahn echos can be canceled when the following resonance condition is met 
\begin{align}
\omega _{L}  t _{2}  
/ 2 \pi \in \mathbb{Z}_{+}
\Rightarrow 
& \, \mathcal{C} _{\mathrm{2he} } 
\left ( t _{1}, t _{2}  
;  \tau \right )
    \left [ 
\hat{\rho}_{\mathrm{b}} \right ] 
\simeq \hat{\rho}_{\mathrm{b}} 
. 
\end{align}
We note that this full cancellation (to leading order in NV-bath coupling) is enabled by the NV spin-$1$ structure.

\end{document}